\documentclass[10pt,dvips,a4paper]{article}
\usepackage{epsfig}

\newcommand{\veps}{\varepsilon}

\newcommand{\ignore}[1]{}

\usepackage{amssymb}
\usepackage{amsmath}
\usepackage{setspace}
\usepackage[numbers]{natbib}

\begin{document}

\renewcommand{\title}{Meso-scale approach to modelling the fracture process zone of concrete subjected to uniaxial tension}

\begin{center} \begin{LARGE} \textbf{\title} \end{LARGE} \end{center}

\begin{center} Peter Grassl$^{1*}$ and Milan Jir\'{a}sek$^{2}$ \end{center}

\begin{center}
$^{1}$Department of Civil Engineering, University of Glasgow, Glasgow, UK\\
Email: grassl@civil.gla.ac.uk\\
Phone: +44 141 330 5208\\
Fax: +44 141 330 4557\\
$^{*}$Corresponding author
\end{center}

\begin{center}
$^{2}$Department of Mechanics, Faculty of Civil Engineering, Czech Technical University in Prague, Czech Republic \\
Email: Milan.Jirasek@fsv.cvut.cz\\
Phone: +420 224 354 481\\
Fax: +420 224 310 775\\
\end{center}

\begin{center}
Submitted to International Journal of Solids and Structures, January 29, 2009
\end{center}

\section{Abstract}

A meso-scale analysis is performed to determine the fracture process zone of concrete subjected to uniaxial tension.
The meso-structure of concrete is idealised as stiff aggregates embedded in a soft matrix and separated by weak interfaces.
The mechanical response of the matrix, the inclusions and the interface between the matrix and the inclusions is modelled by a discrete lattice approach.
The inelastic response of the lattice elements is described by a damage approach, which corresponds to a continuous reduction of the stiffness of the springs.
The fracture process in uniaxial tension is approximated by an analysis of a two-dimensional cell with periodic boundary conditions.
The spatial distribution of dissipated energy density at the meso-scale of concrete is determined.
The size and shape of the deterministic FPZ is obtained as the average of random meso-scale analyses.
Additionally, periodicity of the discretisation is prescribed to avoid influences of the boundaries of the periodic cell on fracture patterns.
The results of these analyses are then used to calibrate an integral-type nonlocal model.

\section{Introduction}
The energy dissipation during the fracture of concrete is influenced by the meso-structure of the material, the loading applied and the geometry of the specimen.
The spatial distribution of the dissipated energy density across the fracture process zone (FPZ) is governed by both statistics and mechanics.
Depending on the type of loading, either statistical or mechanical processes dominate.
For instance, the shape of shear bands in concrete is strongly influenced by the mechanical interaction of aggregates. 
For this type of loading, the deterministic contribution on the shape of the FPZ is significant \citep{KouGeeBre02}.
For tensile fracture the tortuosity of the path of the main crack, which dissipates most of the energy, is predominantly determined by the statistics of the random arrangement of aggregates.
Therefore, the crack paths in concrete subjected to the same loading conditions differ significantly, which implies that a purely deterministic meso-scale model, which does not consider the statistical variation of the fracture paths, cannot describe the FPZ of concrete subjected to tension.
Hence, a direct determination of the mean FPZ by meso-scale analysis requires averaging of the results of meso-scale analyses.
Previous modelling efforts have been focused on an indirect determination of the FPZ by inverse analysis \citep{Car99, BelDubPij03, JirGra04,  IacSluMie06, IacSluMie08}. 
In one of these approaches, strength and fracture energies obtained from specimens of different sizes have been used to determine the width of the FPZ \citep{JirRolGra04}. 
However, the results obtained with this calibration approach were not unique and provided only limited insight into the shape and size of the fracture process of concrete.

The present study aims at determining the FPZ for mode-I fracture of concrete by meso-scale analysis. 
The nonlinear finite element method was used to analyse the spatial distribution of dissipated energy density on the meso-scale of concrete.
The size and shape of the deterministic FPZ was determined as the average of meso-scale analyses.
The results were then used to calibrate a deterministic nonlocal model, which can be used for large-scale structural analysis.
To the authors' knowledge, this type of meso-scale analysis for the determination of the fracture process zone has not been carried out before.

Within the framework of the nonlinear finite element analysis, three main approaches to fracture modeling can be distinguished.
Continuum approaches describe the fracture process by higher-order constitutive models, such as integral-type nonlocal models \citep{BazJir02,GraJir05}.
In continuum models with discontinuities, cracks are described as displacement discontinuities, which are embedded into the continuum description \citep{JirZim01a}.
Finally, discrete approaches describe the nonlinear fracture process as failure of discrete elements, such as trusses and beams \citep{Kaw78,Cun79}.
In recent years, one type of discrete approach based on a lattice determined by Voronoi tesselation has been shown to be suitable for fracture simulations \citep{BolSai98}.
The lattice approach is robust, computationally efficient, and allows for fracture description by a stress-inelastic displacement relationship, similar to continuum models with discontinuities. 
With a specially designed constitutive model, the results obtained with this approach were shown to be mesh-independent \citep{BolSuk05}.
Such a lattice approach is used for the meso-scale modelling in the present study.

In the present work, discrete approaches are divided in the group of particle and lattice models. 
In particle models, the arrangement of particles can evolve, so that neighbours of particles might change during analysis.
Therefore, particle models are suitable to describe processes involving large displacements.
On the other hand, in lattice models the connectivity between nodes is not changed during the analysis, so that contact determination is not required.
Consequently, lattice models are mainly suitable for analysis involving small strains \citep{HerHanRou89, SchMie92b, BolSai98}.

In lattice analyses, the meso-structure of concrete can be described in at least two ways.
In the first approach, the interaction between aggregates is modelled directly by single lattice elements \citep{ZubBaz87}. 
All the nonlinearity of the material response between the aggregates is represented by the stress-strain response at a single point.
This approach is characterised by computational efficiency, since the nodes of the finite element mesh correspond to the centres of concrete aggregates \citep{CusBazCed03b}.
In the second approach, information on the heterogeneous meso-structure of concrete is mapped on a lattice in the form of spatially varying material properties \citep{SchMie92b}.  
This approach requires a finer resolution, since individual aggregates are represented by several lattice elements.
In the present study the latter approach is chosen, since a detailed description of the tortuous crack patterns is of importance.
Because of the fine lattice used, the present study is limited to two-dimensional plane stress analyses with aggregates idealised as cylindrical inclusions.
This oversimplifies the meso-structure of concrete so that a direct comparison with experimental results is difficult.
However, the study may produce qualitative results that are useful for the development of macroscopic nonlocal constitutive models.

The lattices used were generated from randomly placed nodes, which reduced the influence of the alignment on the fracture patterns \citep{JirBaz95,SchGar96}. 
Similar observations have been made for other fracture approaches, in which irregular meshes can reduce the influence of the discretisation on fracture patterns \citep{JirGra08}.
The number of degrees of freedom of the background lattice is reduced by placing lattice elements perpendicular to the boundary of aggregates \citep{BolBer04}.
Small aggregates, which cannot be captured by the background lattice, are approximated by a random field, which is generated by a spectral representation \citep{ShiJan72,ShiDeo96}. 
Thus, the influence of the background lattice on the fracture patterns in the regions between the discretely modelled aggregates is reduced.

The constitutive response of the meso-structure of concrete can either be described by micro-mechanical models based on multi-scale analysis \citep{BudOco76} or, alternatively, by phenomenological constitutive models, which are commonly based on the theory of damage mechanics \citep{Mazars84}, plasticity \citep{Etse94} and combinations of damage and plasticity \citep{Ju89}. 
Plasticity and combinations of damage and plasticity are well suited to describe the fracture process of concrete in compression \citep{GraJir06, GraLunGyl02}.
On the other hand, for mode-I fracture, damage constitutive models often provide satisfactory results \citep{JirGra08}.  
In the present study an isotropic damage constitutive model is used.

\section{Meso-scale modelling approach}

The present approach to modelling the fracture process zone of concrete is based on a meso-scale description. Aggregates, 
interfacial transition zones (ITZ) and mortar are modelled as separate phases with different material properties. 
For the mortar and ITZ, a random field of strength and fracture energy is applied.
A lattice approach \citep{BolSai98} is used in combination with a damage mechanics model.

\subsection{Lattice approach} 
\begin{figure}
\begin{center}
\begin{tabular}{cc}
\epsfig{file=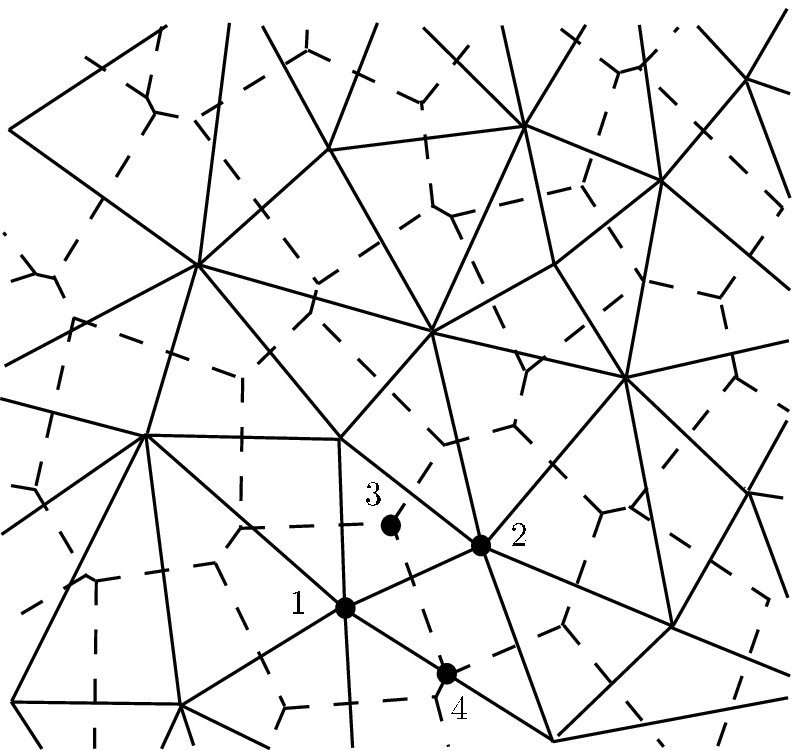,width=8cm} & \epsfig{file=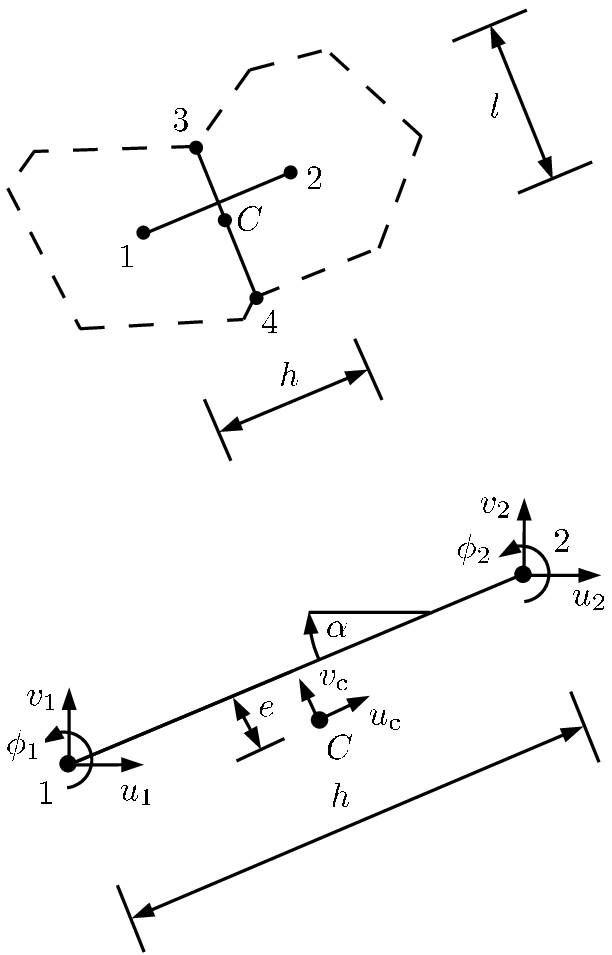,width=6cm} \\
(a) & (b)
\end{tabular}
\end{center}
\caption{(a) Lattice based on Voronoi polygons. (b) Lattice element in the global coordinate system.}
\label{fig:lattice}
\end{figure}

The material response is represented by a discrete system of structural elements.
The nodes of the lattice are randomly located in the domain, subject to the constraint of a minimum distance $d_{\rm min}$ \citep{ZubBaz87}.
For the discretely modelled aggregates, the lattice nodes are not placed randomly but at special locations,
such that the middle cross-sections of the lattice elements form the boundaries between aggregates and mortar \citep{BolBer04}.
The lattice elements are obtained from the edges of the triangles of the Delaunay triangulation of the domain (Fig.~\ref{fig:lattice}a), whereby the middle cross-sections of the lattice elements are the edges of the polygons of the dual Voronoi tesselation \citep{BolSai98}.

Each lattice node possesses three degrees of freedom (two translations and one rotation). 
In the global coordinate system, shown in Figure~\ref{fig:lattice}b, the degrees of freedom $\mathbf{u}_{\rm e} = (u_1, v_1, \phi_1, u_2, v_2, \phi_2)^{\rm T}$ of the lattice nodes are linked to the displacement discontinuities $\mathbf{u}_{\rm c} = (u_{\rm c}, v_{\rm c})^{\rm T}$ in the local co-ordinate system at point $C$, which is located at the center of the middle cross-section of the  element.
The relation between the degrees of freedom and the displacement discontinuities at $C$ is 
\begin{equation}\label{eq1}
\mathbf{u}_{\rm c} = \mathbf{B} \mathbf{u}_{\rm e}
\end{equation}
where
\begin{equation}
\mathbf{B} = \begin{bmatrix}
-\cos \alpha & - \sin \alpha & -e & \cos \alpha & \sin \alpha & e\\
\sin \alpha & - \cos \alpha & -h/2 & \cos \alpha & \sin \alpha & -h/2
\end{bmatrix}
\end{equation}

The displacement discontinuities $\mathbf{u}_{\rm c}$ at point $C$ are transformed into strains $\boldsymbol{\veps} = (\veps_{\rm n}, \veps_{\rm s})^{T} = \mathbf{u}_{\rm c} /h$, where $h$ is the distance between the two lattice nodes. 
The strains 
are related to the stresses $\boldsymbol{\sigma} = (\sigma_{\rm n}, \sigma_{\rm s})^{T}$ by an isotropic damage model, which is described below.

The stiffness matrix of the lattice element in the local coordinate system has the form
\begin{equation}
\mathbf{K} = \dfrac{l}{h} \mathbf{B}^{\rm T} \mathbf{D} \mathbf{B}
\end{equation}
where $l$ is the cross-sectional area 
(for a two-dimensional model this area reduces 
to the length of the side shared by neighboring Voronoi polygons)
and $\mathbf{D}$ is the material stiffness matrix.

Heterogeneous materials are characterised by spatially varying material properties.
In the present work this is reflected at two levels.
Aggregates with diameters greater than $\phi_{\rm min}$  are modelled directly.
The random distribution of the aggregate diameters $\phi$ is defined by the cumulative distribution function used in \citep{CarCorPuz04, GraRem08}. 
The aggregates are placed randomly within the area of the specimen, avoiding overlap of aggregates. 
Overlaps with boundaries are permitted.
The heterogeneity represented by finer particles is described by an autocorrelated random field of tensile strength and fracture energy, which are assumed to be fully correlated.
The random field is generated using a spectral representation \citep{ShiJan72,ShiDeo96}.
This mixed approach, in the form of a discrete representation of the meso-structure and random field, is a compromise between model detail and computational time.

The random field is characterised by an exponential autocorrelation function 
\begin{equation}
R\left(\xi\right) = \exp\left(-|\xi|^2/b^2\right)
\end{equation}
and a Gaussian probability distribution function with a threshold to exclude negative values of strength and fracture energy.
Parameter $\xi$ is the separation distance and parameter $b$ is related to the autocorrelation length $l_{\rm a}$ as 
\begin{equation}
b = \dfrac{2 l_{\rm a}}{\sqrt{\pi}}
\end{equation}
The autocorrelation length $l_{\rm a}$ is independent of the spacing $d_{\rm min}$ of the lattice nodes \citep{GraBaz09}.

\subsection{Constitutive model}\label{sec:const}
An isotropic damage model is used to describe the constitutive response of ITZ and mortar.
In the following section, the main equations of the constitutive model are presented. 
The stress-strain law reads
\begin{equation} \label{eq:totStressStrain}
\boldsymbol{\sigma} = \left(1-\omega \right) \mathbf{D}_{\rm e} \boldsymbol{\varepsilon}  = \left(1-\omega\right) \bar{\boldsymbol{\sigma}}
\end{equation}
where $\boldsymbol{\sigma} = \left(\sigma_{\rm n}, \sigma_{\rm s}\right)^T$ is the nominal stress,
$\omega$ is the damage variable, 
$\mathbf{D}_{\rm e}$ is the elastic stiffness, 
$\boldsymbol{\varepsilon} = \left(\varepsilon_{\rm n},\varepsilon_{\rm s}\right)^T$ is the strain,
and $\bar{\boldsymbol{\sigma}} = \left(\bar{\sigma}_{\rm n}, \bar{\sigma}_{\rm s}\right)^T$ is the effective stress.

The elastic stiffness 
\begin{equation}
\mathbf{D}_{\rm e} = \begin{bmatrix} E & 0\\
0 & \gamma E
\end{bmatrix}
\end{equation}
depends on model parameters
$E$ and $\gamma$, which  control Young's modulus and Poisson's ratio of the material \citep{GriMus01}.
For a regular lattice and plane stress, Poisson's ratio is
\begin{equation}
\nu = \dfrac{1-\gamma}{3+\gamma}
\end{equation}
and the elastic stiffness is
\begin{equation}
E_{\rm m} = 2 E \left(\dfrac{1+\gamma}{3+\gamma}\right)
\end{equation}

For a positive shear stiffness, i.e.\ $\gamma >0$, Poisson's ratio is limited to $\nu<\dfrac{1}{3}$, which is acceptable for concrete but may be unrealistic for certain other materials.
The damage parameter $\omega$ is a function of a history variable $\kappa$, which is determined by the loading function
\begin{equation}
f(\boldsymbol{\varepsilon},\kappa) = \varepsilon_{\rm eq} \left( \boldsymbol{\varepsilon} \right) - \kappa 
\end{equation}
and the loading-unloading conditions 
\begin{equation}\label{loadunload}
f \leq 0 \mbox{,} \hspace{0.5cm} \dot{\kappa} \geq 0 \mbox{,} \hspace{0.5cm} \dot{\kappa} f = 0
\end{equation}

\begin{figure}
\begin{center}
\epsfig{file=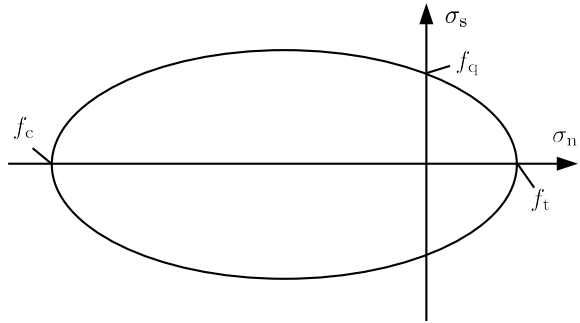, width=8cm}
\caption{Elliptic stress envelope in the nominal stress space.}
\label{fig:envelope}
\end{center}
\end{figure}

The equivalent strain 
\begin{equation} \label{eq:equiv}
\varepsilon_{\rm eq}(\varepsilon_{\rm n},\varepsilon_{\rm s}) = \dfrac{1}{2} \varepsilon_0 \left( 1-c \right) + \sqrt{\left( \dfrac{1}{2} \varepsilon_0 (c-1) + \varepsilon_{\rm n}\right)^2 + \dfrac{ c \gamma^2 \varepsilon_{\rm s}^2}{q^2}}
\end{equation} 
corresponds to an elliptic stress envelope in the nominal stress space (Fig.~\ref{fig:envelope}).

For pure tensile loading, the nominal stress is limited by the tensile strength $f_{\rm t} = E \varepsilon_{0}$.
For pure shear and pure compression, the nominal stress is limited by the shear strength $f_{\rm q} = q f_{\rm t}$ and the compressive strength $f_{\rm c} = c f_{\rm t}$, respectively.

The softening curve of the stress-strain response in uniaxial tension is defined by the relation
\begin{equation} \label{eq:exp}
\sigma = f_{\rm t} \exp \left(-\dfrac{w_{\rm c}}{w_{\rm f}}\right)
\end{equation}
where $w_{\rm c} = \omega h \varepsilon$ is considered as an equivalent crack opening under monotonic uniaxial tension. 
The stress-strain law in uniaxial tension can also be expressed in terms of the damage variable as
\begin{equation}\label{eq:uni}
\sigma =  \left(1-\omega \right) E \varepsilon
\end{equation}
Comparing the right-hand sides of (\ref{eq:exp}) and (\ref{eq:uni}), 
and replacing $\varepsilon$ by $\kappa$, we obtain the equation
\begin{equation}
\left(1-\omega \right) \kappa = \varepsilon_0\exp \left(-\dfrac{\omega h \kappa}{w_{\rm f}}\right)
\end{equation}
from which the damage parameter $\omega$ can be determined iteratively.
Parameter $w_{\rm f}$ determines the initial slope of the softening curve and is related to the meso-level
fracture energy $G_{\rm ft} = f_{\rm t} w_{\rm f}$.
The present damage constitutive model used here is conceptually similar to  \citep{OrtPan99}.
However, a more complex equivalent strain formulation involving the additional parameter $c$ was required in the present study, since the compressive strength of concrete is significantly greater than its tensile strength.
For $c=1$ the original formulation proposed in \citep{OrtPan99} is obtained.
\subsection{Periodic cell}

One important issue in micro- and meso-mechanical modeling is the specific choice of boundary conditions
at the boundary of the computational cell. 
In elastic homogenization theory, it is well known that kinematic boundary conditions (imposed displacements)
lead to an apparent stiffness (describing the relation between average strain and average stress) that
overestimates the effective one,
while static boundary conditions (imposed tractions) lead to an underestimated stiffness. 
The influence of boundary conditions on the apparent elastic properties
 fades away with increasing size of the cell.
If the cell is large enough, it is considered as a representative volume element and the apparent elastic properties
are close to the effective properties of the homogenized medium. 

If the analysis is extended to highly nonlinear material behavior that leads to localization of inelastic processes,
the boundary conditions may have a strong influence even if the cell is very large. The reason is that the 
imposed constraints usually lead to stress concentrations in a material layer near the boundary, which then bias
the localization pattern. It is therefore preferable to use boundary conditions that are free of such bias
and provide a statistically homogeneous distribution of localization patterns. The undesired boundary effects
can be eliminated by the periodicity requirement that replaces the boundary conditions. This requires a special
meso-structure and its discretization (in our case lattice), both compatible with the periodicity requirement. 

The meso-scale approach used in this paper is based on the concept of an initially rectangular 
computational cell that can be periodically
repeated in the directions parallel to its sides, both in the initial and the current (deformed) configuration.
In addition to the displacements and rotations of individual lattice nodes that
are contained withing the cell, the average strain components are considered as degrees of freedom of the
computational model. Force quantities work-conjugate to these global degrees of freedom are directly related
to the average stress components. It is thus easy to impose various types of mixed loading conditions.
For instance, for uniaxial tension at the macroscopic scale, the average strain in the direction of loading
is incremented (playing the role of the control variable) 
while the average stress in the lateral direction and the average shear stress are set to zero.

\begin{figure}
\begin{center}
\epsfig{file=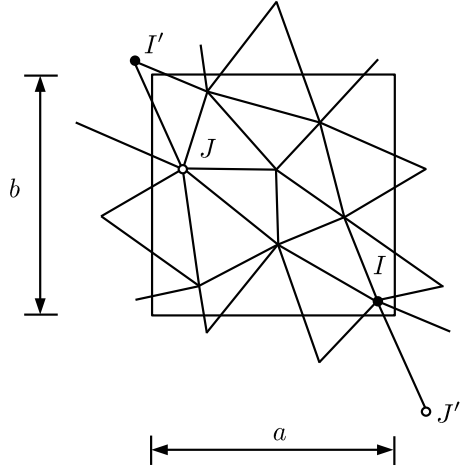,width=6cm}
\caption{Periodic cell.}
\label{fig:periodicCell}
\end{center}
\end{figure}

When setting up the discretised equilibrium equations of the meso-scale model, the internal lattice elements,
i.e.\ those connecting
two nodes that are both inside the cell, are processed in the standard way. 
Special treatment is needed for the boundary elements that connect one node inside the cell with another node
physically located
in one of the neighboring cells. The external lattice node is a periodic image of one of the internal nodes.
For easier reference, let us denote $I$ and $J$ the internal lattice nodes and $I'$ and $J'$ their periodic
images outside the cell; see Fig.~\ref{fig:periodicCell}. 

The corresponding lattice element can be considered
as connecting either nodes $I$ and $J'$, or $I'$ and $J$. Both representations are equivalent and only one of them,
say  $I$ and $J'$,
is actually processed when setting up the contribution to the equilibrium equations. Due to the assumption
of periodicity, the rotations of lattice nodes $J$ and $J'$ are the same and the translations of node $J'$
can be expressed as  the translations of node $J$ plus an appropriate contribution of the average strains.
More specifically, suppose that the initial position of node $J'$ is shifted with respect to node $J$ by $k_x a$ in the
$x$-direction and by $k_y b$ in the $y$-direction, where $a$ and $b$ are dimensions of the cell and $k_x$ and $k_y$
are integers $-1$, 0 or 1 (for the specific case plotted in Fig.~\ref{fig:periodicCell}, we have $k_x=1$ and $k_y=0$).
The translations of node $J'$ are then expressed as
\begin{eqnarray}
\label{period1}
u_{J'} &=& u_J + k_x a E_x + k_y b E_{xy} \\
\label{period2}
v_{J'} &=& v_J + k_y b E_y 
\end{eqnarray}
where $E_x$ and $E_y$ are the average normal strains in the $x$- and $y$-direction, respectively,
and $E_{xy}$ is the average engineering shear strain. Note that the contribution of the average shear
has been attributed exclusively to the translations along the $x$-direction. This is perfectly justified,
because rigid-body rotation of the entire lattice is completely arbitrary and we can fix it by assuming
that straight  lines parallel to the $x$-axis do not rotate. The other two rigid-body modes (translations)
need to be suppressed by setting the translations of one selected lattice node to zero. 

Making use of (\ref{period1})--(\ref{period2}) and of the relation for the rotations, 
$\phi_{J'}=\phi_J$, we can set up the 
transformation rule 
\begin{equation}\label{transf}
\left(\begin{array}{c}
u_I \\ v_I \\ \phi_I \\ u_{J'} \\ v_{J'} \\ \phi_{J'}
\end{array}\right)
=
\left[\begin{array}{ccccccccc}
1 & 0 & 0 & 0 & 0 & 0 & 0 & 0 & 0 \\ 
0 & 1 & 0 & 0 & 0 & 0 & 0 & 0 & 0 \\ 
0 & 0 & 1 & 0 & 0 & 0 & 0 & 0 & 0 \\ 
0 & 0 & 0 & 1 & 0 & 0 & k_x a & 0 & k_y b \\ 
0 & 0 & 0 & 0 & 1 & 0 & 0 & k_y b & 0 \\ 
0 & 0 & 0 & 0 & 0 & 1 & 0 & 0 & 0 
\end{array}\right]
\left(\begin{array}{c}
u_I \\ v_I \\ \phi_I \\ u_J \\ v_J \\ \phi_J \\ E_x \\ E_y \\ E_{xy}
\end{array}\right)
\end{equation}
When this rule is combined with equation (\ref{eq1})
linking the displacements of nodes $I$ and $J'$ and the discontinuities, 
the $6\times 9$ transformation matrix from  (\ref{transf}), denoted as $\mathbf{T}$,
multiplies matrix $\mathbf{B}$ from the right. It follows from duality that
the internal forces must be multiplied by $\mathbf{T}^T$ from the left before
they are inserted into the equilibrium conditions. The usual $6 \times 6$ stiffness matrix
of the element, $\mathbf{K}$, is therefore transformed into a $9 \times 9$ matrix  
$\mathbf{T}^T\mathbf{K}\mathbf{T}$ before it enters the assembly process. 
The global stiffness matrix remains banded, with the exception of the 
three rows and columns that contain terms related to the global degrees of
freedom (average strains) and to the corresponding equilibrium equations
(related to the average stresses).

\section{Meso-scale analysis}

\subsection{Geometry, loading setup and results}

The aim of the present study is to analyse the fracture process zone of concrete under uniaxial tension.
A specimen with periodic boundary conditions, material properties and background lattice was used to reduce the influence of boundaries on the fracture process.
The material properties for the mortar, ITZ and aggregate phases were chosen according to Tab.~\ref{tab:param}.
\begin{table}
\begin{center}
\caption{Model parameters.}
\label{tab:param}
\begin{tabular}{cccccccc}
\hline \\
 & $E$ [GPa] & $\gamma$ & $\bar{f}_{\rm t}$~[MPa] & $q$ & $c$ & $\bar{G}_{\rm ft}$~[N/m]\\
Matrix & $30$ & $0.33$ & $5.3$ & $2$ & $10$ & $93$\\
ITZ & $45$ & $0.33$ & $1.8$ & $2$ & $10$  & $31$\\
Aggregate & $90$ & $0.33$ & - & - & - & -\\
\hline\\
\end{tabular}
\end{center}
\end{table}
Parameters $\bar{f}_{\rm t}$ and $\bar{G}_{\rm f}$ in Tab.~\ref{tab:param} are the mean values of the random fields of tensile strength $f_{\rm t}$ and fracture energy $G_{\rm f}$, respectively.
The material parameters were chosen to approximate a macroscopic tensile strength of 3~MPa and a macroscopic fracture energy of 100 J/m$^2$.
The ratio of the Young's modulus of matrix and aggregate is assumed to be 3. The Young's modulus of the ITZ is 
\begin{equation}
E_{\rm{ITZ}} = \dfrac{2}{\dfrac{1}{E_{\rm m}} + \dfrac{1}{E_{\rm a}}}
\end{equation}
where $E_{\rm m}$ and $E_{\rm a}$ are the Young's moduli of matrix and aggregate, respectively.
Furthermore, the ratios of tensile strength and fracture energy for matrix and ITZ are 3.
These values are in the range used for other meso-scale analyses in \citep{GraRem08,RemGra08}.

\begin{figure}
\begin{center}
\begin{tabular}{cc}
\epsfig{file=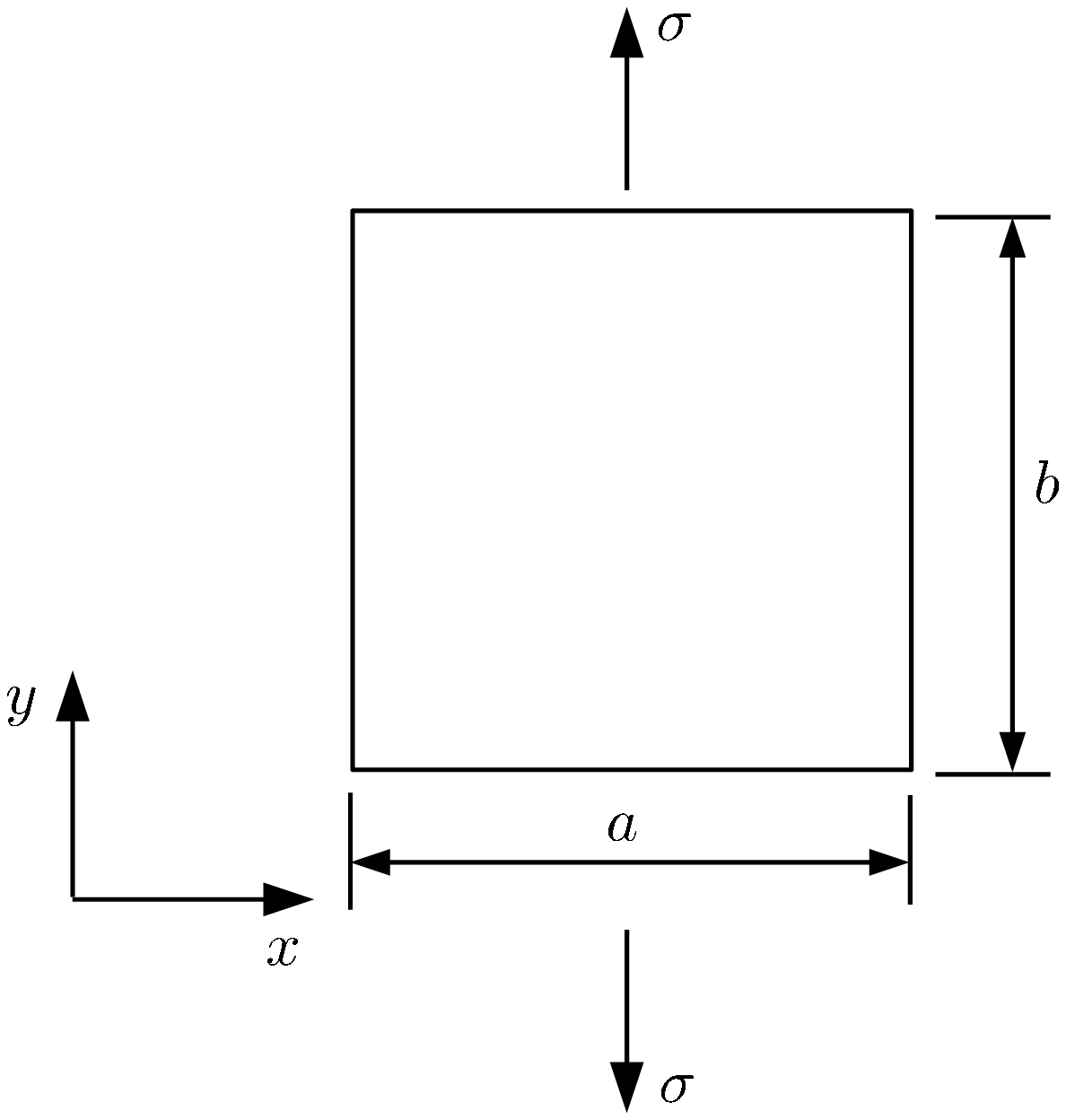,width=6cm} & \epsfig{file=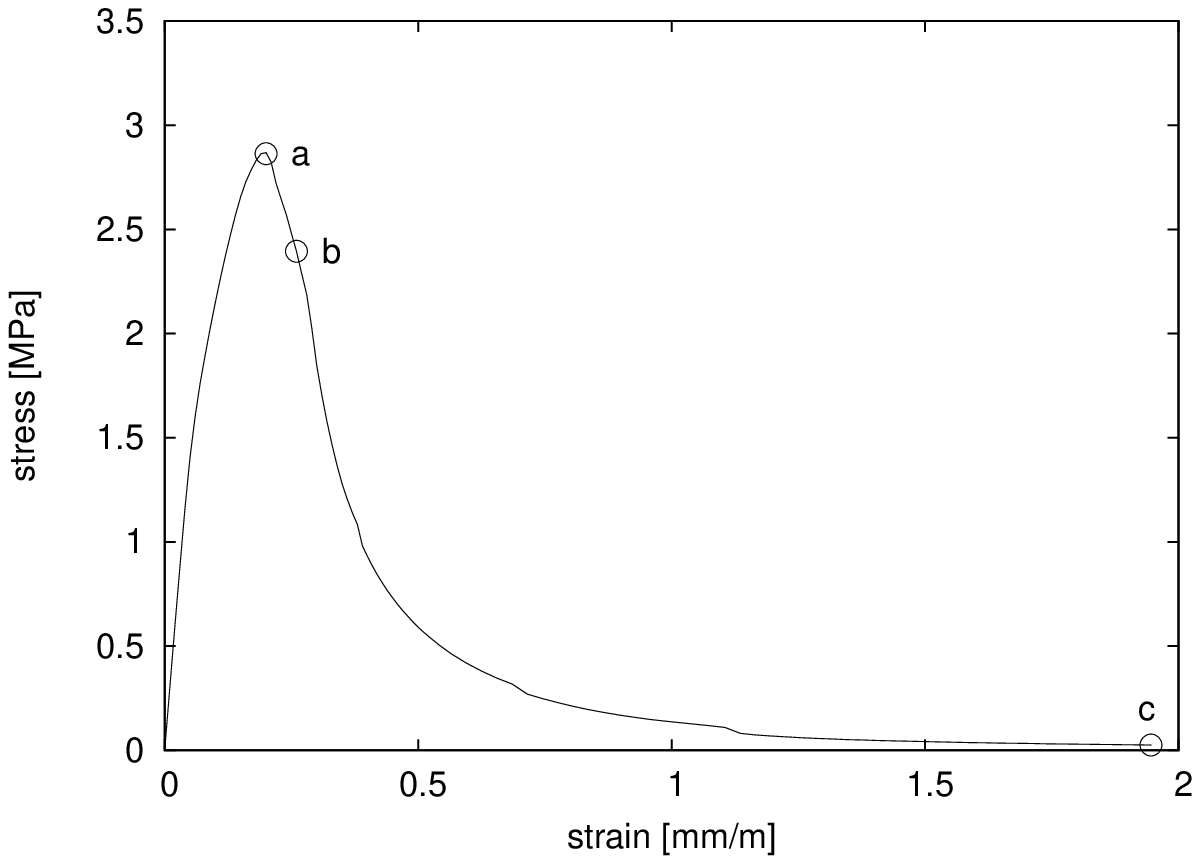,width=10cm} \\
(a) & (b)
\end{tabular}
\end{center}
\caption{(a) Geometry and loading setup of the periodic specimen with $a = b = 100$~mm. (b) Load-displacement curve for one random meso-scale analysis.}
\label{fig:geometry}
\end{figure}

The specimen used in the analyses (Fig.~\ref{fig:geometry}) had a height and width of $a = b = 100$~mm, respectively.
The lattice for the meso-scale analysis was generated with $d_{\rm min} = 0.75$~mm.
The aggregate volume fraction was chosen as $\rho=0.3$ with a maximum and minimum aggregate diameter of $\phi_{\rm max}=12$~mm and $\phi_{\rm min}=4.75$~mm, respectively. The approach to generate the distribution of aggregate diameters is described in \citep{GraRem08}.
Furthermore, the random field was characterised by the autocorrelation length $l_{\rm a} = 1$~mm and the coefficient of variation $c_{\rm v} = 0.2$.
The specimen was subjected to uniaxial tensile stress $\sigma_{\rm y}$, which was controlled by the displacement in y-direction.
On the left and right boundary of the specimen, boundary conditions that correspond to a vanishing macroscopic stress in the $x$-direction were applied.

\begin{figure}
\begin{center}
\begin{tabular}{ccc}
\epsfig{file=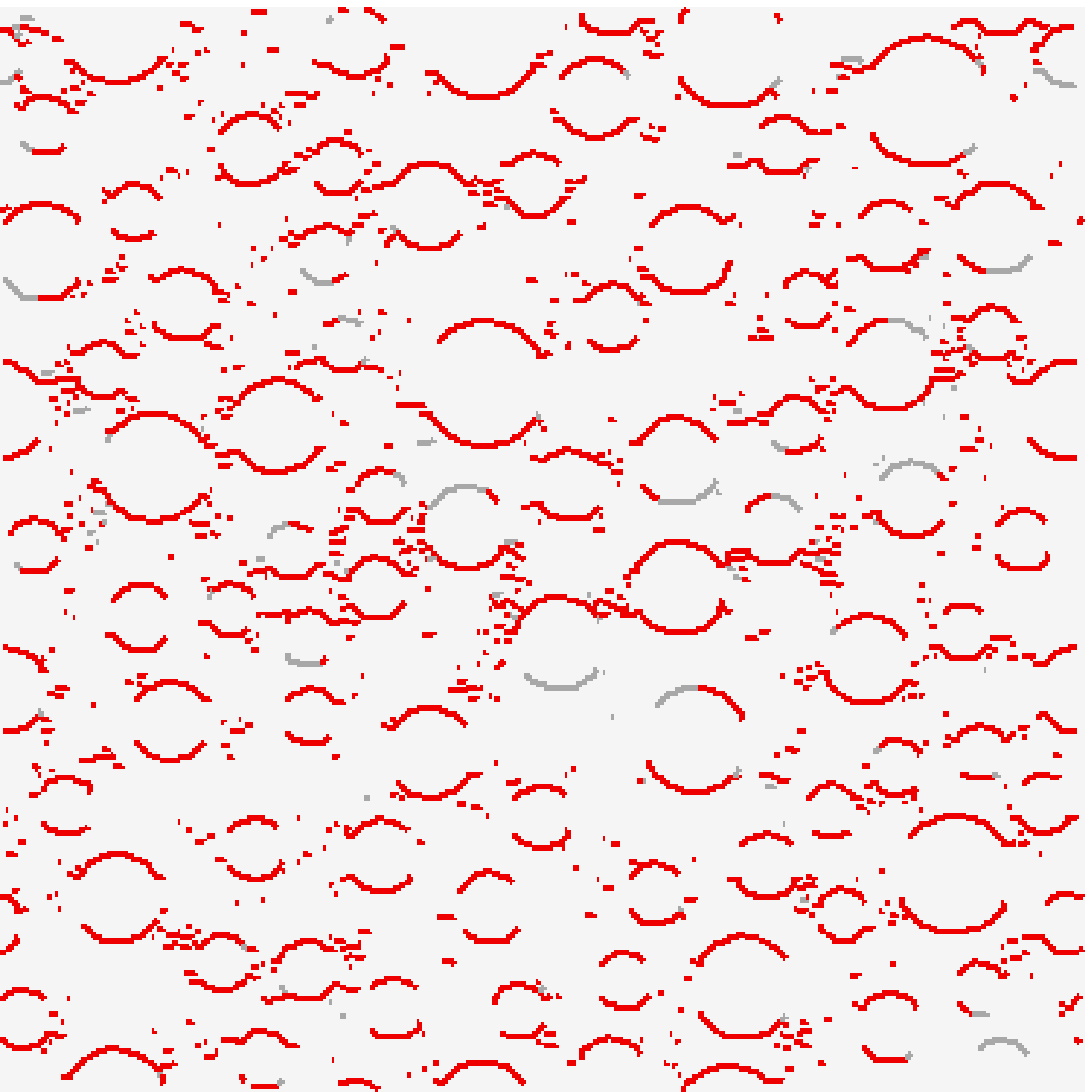,width=4cm} & \epsfig{file=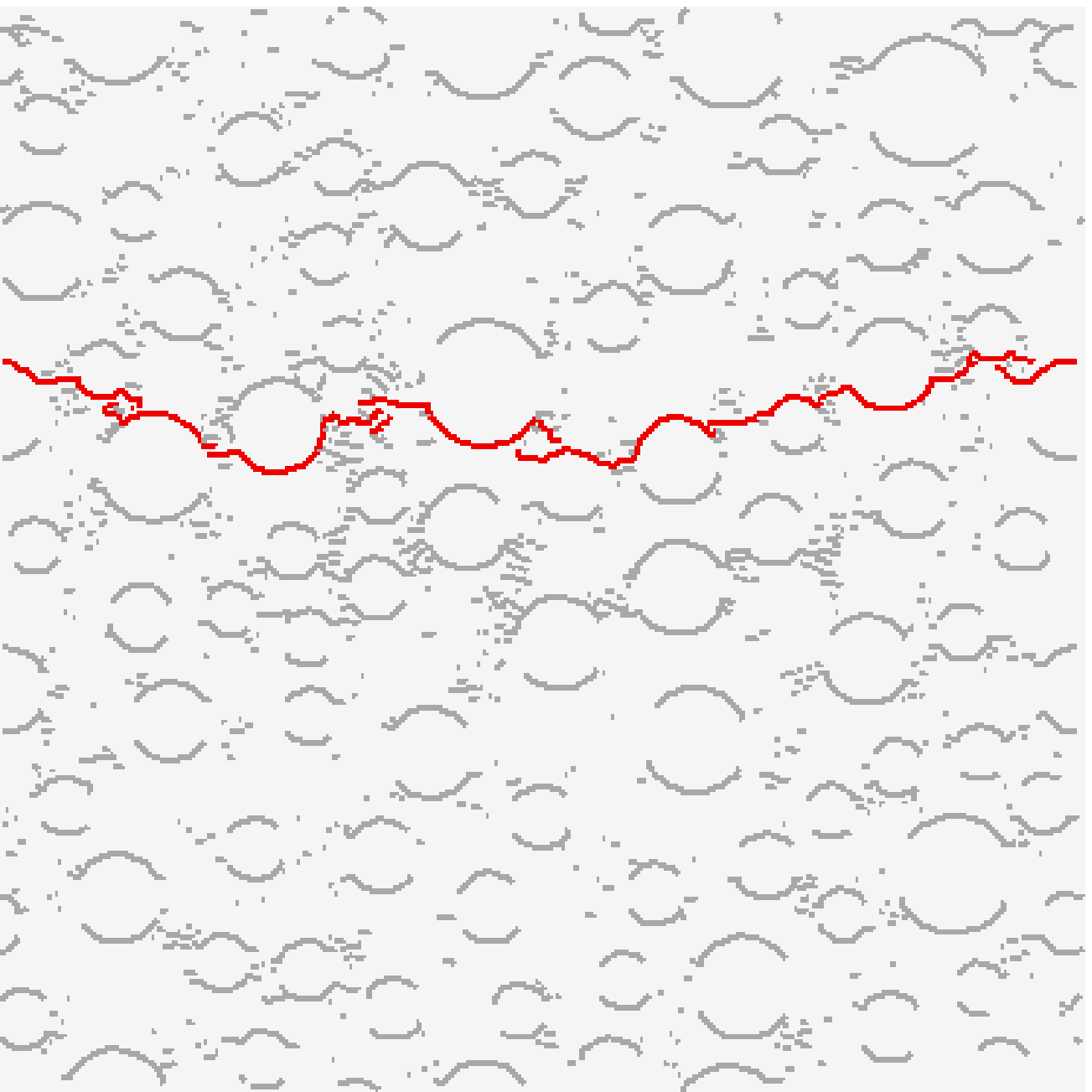,width=4cm} & \epsfig{file=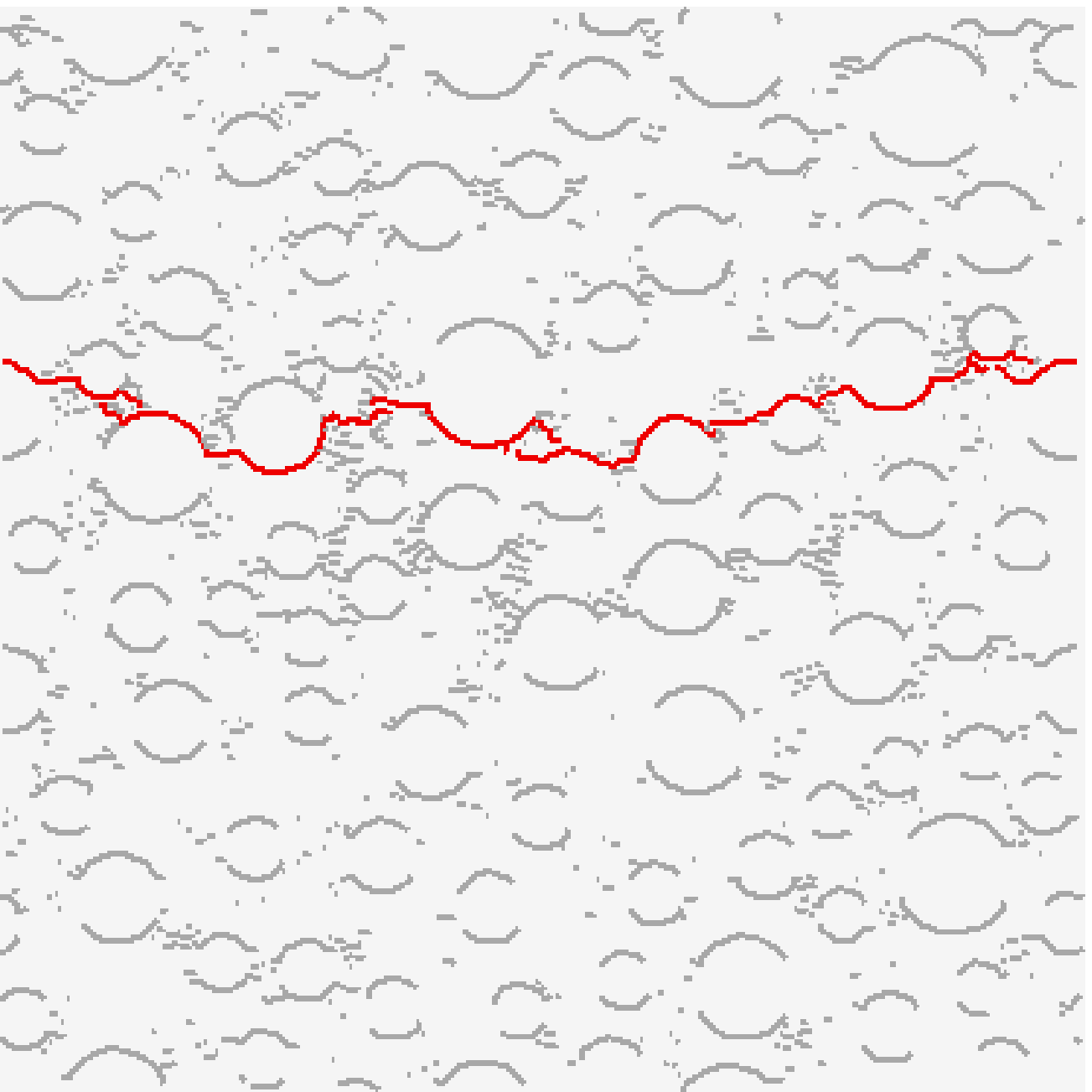,width=4cm} \\
(a) & (b) & (c)
\end{tabular}
\end{center}
\caption{Crack patterns for three stages of loading marked in Fig.~\protect \ref{fig:geometry}b. Red (dark grey) lines indicate cross-sections with increasing crack opening. Light grey lines indicate cross-sections with decreasing crack openings.}
\label{fig:cracks}
\end{figure}

\begin{figure}
\begin{center}
\epsfig{file=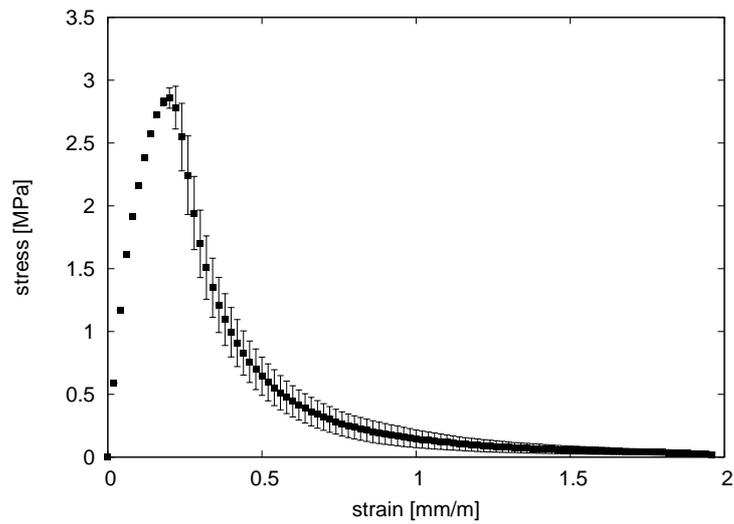,width=10cm}
\end{center}
\caption{Mean curve obtained by averaging the results of 100 analyses. Error bars show the range between the mean plus and minus one 
standard deviation.}
\label{fig:ld}
\end{figure}

The stress-strain response for one random meso-scale analysis is shown in Fig.~\ref{fig:geometry}b.
The crack patterns for three stages of the uniaxial tensile loading, marked in Fig.~\ref{fig:geometry}b, are shown in Fig.~\ref{fig:cracks}. Red (dark grey) lines mark cross-sections of elements in which the damage parameter increases at this stage of analysis (i.e., active, opening cracks), whereas light grey lines indicate cross-sections in which the maximum damage parameter was reached at an earlier stage of analysis (i.e., passive, closing cracks).
From Fig.~\ref{fig:geometry}b~and~\ref{fig:cracks} it can be seen that the path of the active crack is highly tortuous. At earlier stages of analysis, distributed cracking occurs, whereby the cracks are located mainly in the ITZs (light grey cracks) (Fig.~\ref{fig:cracks}~a). Shortly after peak (Fig.~\ref{fig:cracks}~b), many of these cracks cease to grow, i.e. their opening decreases, and only few cross-sections exhibit increasing crack opening (red or dark grey lines).
This is in agreement with observations made in \citep{CedDeiIor87,PlaEliGui92,NirHor92,BolHikShi93}.
Energy dissipation in these localised cracks constitutes a substantial part of dissipated energy.
In Fig.~\ref{fig:ld}, the mean of 100 meso-scale analyses is presented with error bars showing the standard deviation. The results are plotted in terms of the average stress (force per unit length of the unit cell side and per unit thickness) against the average strain (relative displacement of the opposite sides divided by their distance).  
The average stress-strain curve exhibits pre-peak nonlinearities typical of uniaxial tensile experiments of concrete.
The standard deviation is small in the pre-peak regime, but increases strongly in the post-peak regime of the analysis.

\begin{figure}
\begin{center}
\epsfig{file=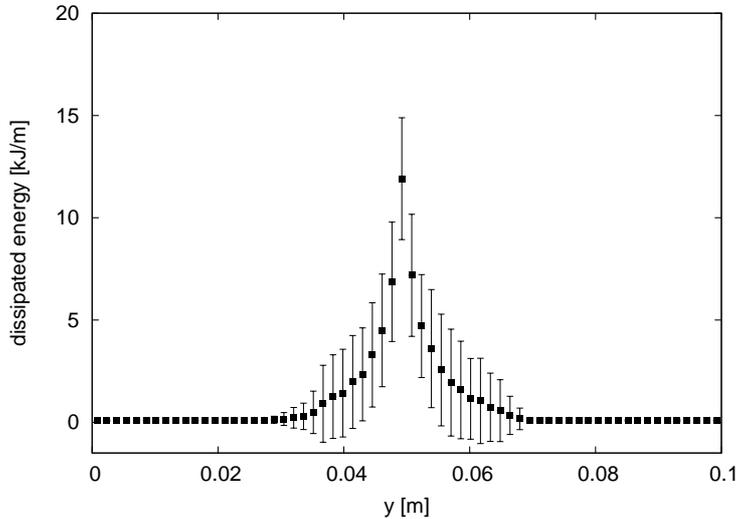,width=10cm}
\end{center}
\caption{Mean profile of the dissipated energy density across the process zone obtained by averaging 100 analyses with a resolution of $1.6$~mm. Error bars show the range between the mean plus and minus one
standard deviation. Fracture process zones are shifted before averaging.}
\label{fig:fpzAverage}
\end{figure}

\begin{figure}
\begin{center}
\epsfig{file=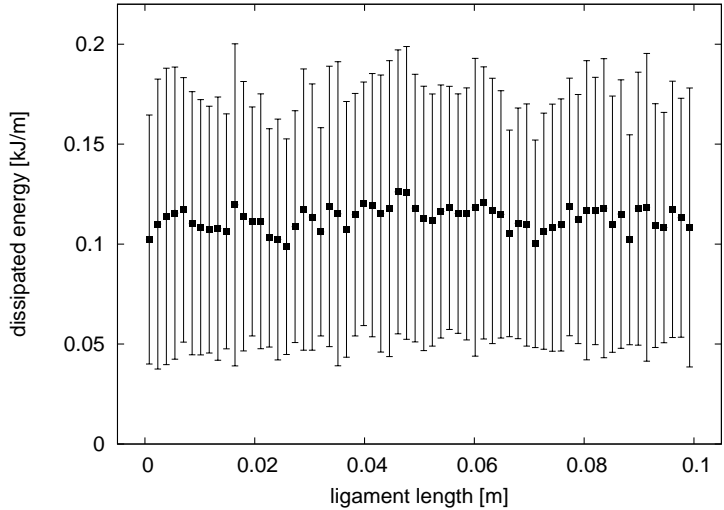,width=10cm}
\end{center}
\caption{Mean values of the dissipated energy density along the process zone obtained by averaging 100 analyses with a resolution of $1.6$~mm. Error bars show the range between the mean plus and minus one
standard deviation.}
\label{fig:rCurve}
\end{figure}

In addition to the stress-strain curves in Fig.~\ref{fig:ld}, the dissipated energy densities are evaluated. 
The averaging of dissipated energy densities is complicated, since the location of the final crack, in which most of the energy is dissipated, depends on the meso-structure of the material and cannot be determined in advance. 
Direct averaging of the energy densities for random meso-scale analyses, would lead, in the limit of an infinite number of analyses, to a uniformly distributed energy density over the specimen height.
For a meaningful characterization of the random fracture process zones, the results of individual analysis were post-processed. 
Firstly, the $y$-coordinate of the centre of the fracture process was determined by processing the dissipated energy of all elements.
Then, all elements were shifted in the $y$-direction (Fig.~\ref{fig:geometry}), such that the $y$-coordinate of the centre of the dissipated energy density of each individual analysis coincides with the $y$-coordinate of the centre of the specimen.
This shift is admissible, since the specimen is fully periodic, i.e.\ not only the boundary conditions, but also meso-structure and background lattice are periodic.
In the next step, the specimen was subdivided in a regular rectangular grid of cells with $x$-~and~$y$-edge lengths of $c_{\rm x} = a/64$ and $c_{\rm y} = b/64$. 
The energy densities in these cells were determined by integrating the dissipated energy of all elements located within them, and by dividing these energies by the cells areas.
Subsequent averaging of the energy densities of the 100 analyses results in the average dissipated energy density, which characterizes the fracture process zone.
The energy density along the $y$-direction, i.e.\ perpendicular to the crack, is presented in Fig.~\ref{fig:fpzAverage} by averaging the dissipated energy along the crack (along the $x$-direction). The error bars in Fig.~\ref{fig:fpzAverage} represent the range between the mean plus and minus one standard deviation.
The fracture process zone has its maximum value in the centre with symmetrically decreasing slopes on both sides. 
The width of the fracture process zone is mainly determined by the tortuosity of the cracks and is for the present meso-structure equal to approximately. 3 times the size of the maximum aggregates.
The dissipated energy density in the $x$-direction (along the crack) is shown in Fig.~\ref{fig:rCurve}.
For this representation, the density is integrated in the $y$-direction.
Thus, the result corresponds to the energy dissipated per unit area of the ligament (per unit length in the two-dimensional setting).
The distribution of dissipated energy along the crack in Fig.~\ref{fig:rCurve} is fluctuating around a constant mean value, which corresponds to the fracture energy of the material. 
Again, error bars indicate the mean plus and minus one standard deviation.


\subsection{Discussion of the meso-scale analysis results}
\begin{figure}
\begin{center}
\epsfig{file=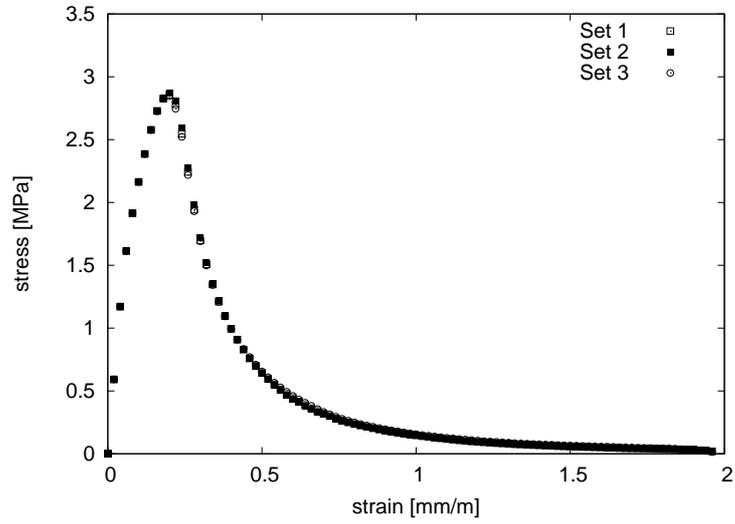,width=10cm}
\end{center}
\caption{Comparison of average stress-strain curves obtained from three different sets of 100 analyses.}
\label{fig:ldSets}
\end{figure}

\begin{figure}
\begin{center}
\epsfig{file=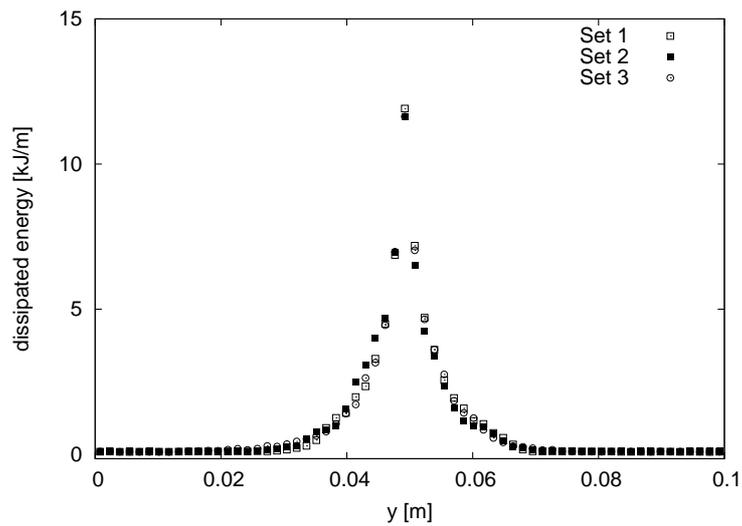,width=10cm}
\end{center}
\caption{Comparison of average fracture process zones obtained from three different sets of 100 analyses. Fracture processes zones are shifted before averaging.}
\label{fig:fpzAverageSets}
\end{figure}

The standard deviations of the dissipated energy densities are considerably greater than those of the stress-strain curves.
At some points of the fracture process zone in Fig.~\ref{fig:fpzAverage}, the standard deviation is even greater than the mean value, which indicates that the statistical distribution is far from normal (since all dissipation values are non-negative). 
The difference in standard deviation is expected, since the average stress-strain curves are global results, which involve averaging of local responses, whereas the dissipated energy density is local.
With such significant standard deviations, it is necessary to check whether 100 analyses are sufficient to obtain statistically representative results.
Therefore, two additional sets of 100 analyses were carried out.
The averages of the three sets are presented in Fig.~\ref{fig:ldSets} and \ref{fig:fpzAverageSets} in the form of stress-strain curves and fracture process zones, respectively.
The difference between the results of the three sets is small, which indicates that 100 analyses are sufficient to obtain statistically representative results.

\begin{figure}
\begin{center}
\epsfig{file=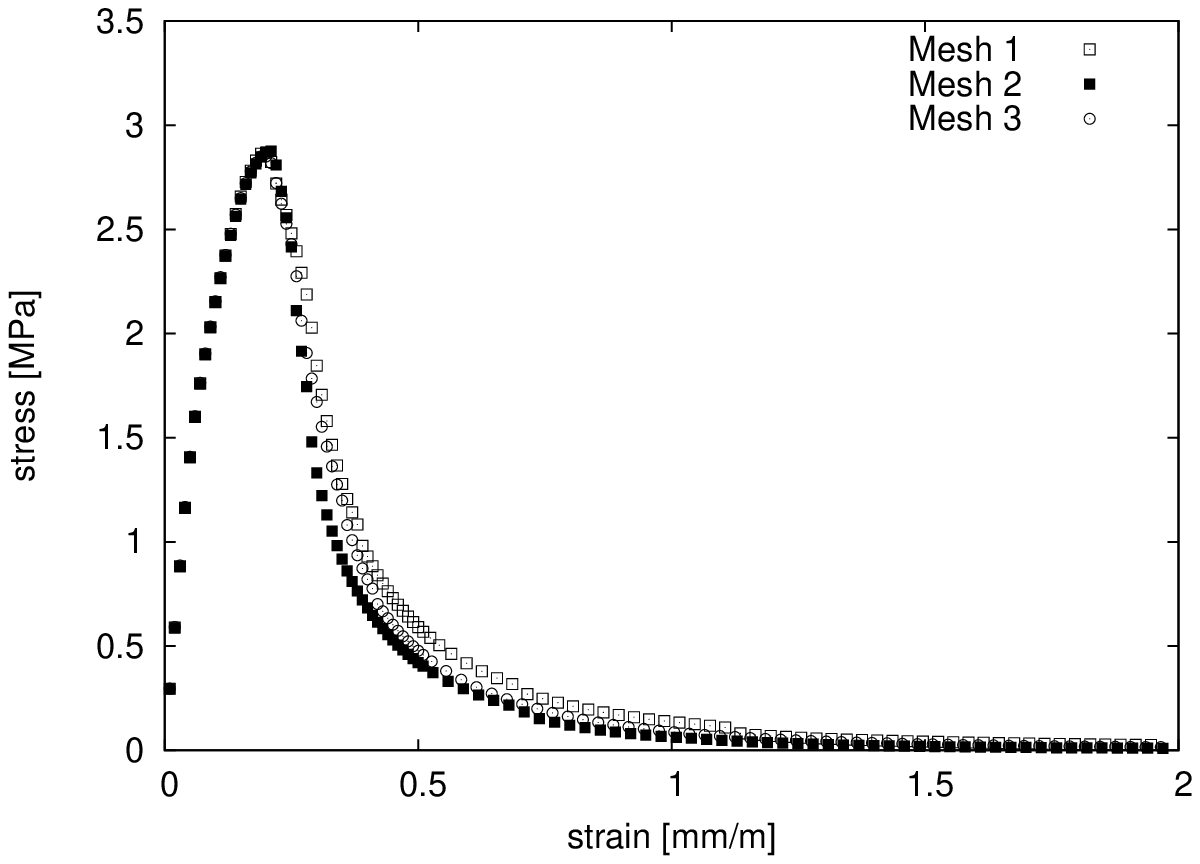,width=10cm}
\end{center}
\caption{Comparison of average stress-strain curves obtained using three different discretisations with the same meso-structure and lattice element size, but different random positions of lattice nodes.}
\label{fig:ldRandom}
\end{figure}

\begin{figure}
\begin{center}
\begin{tabular}{ccc}
\epsfig{file=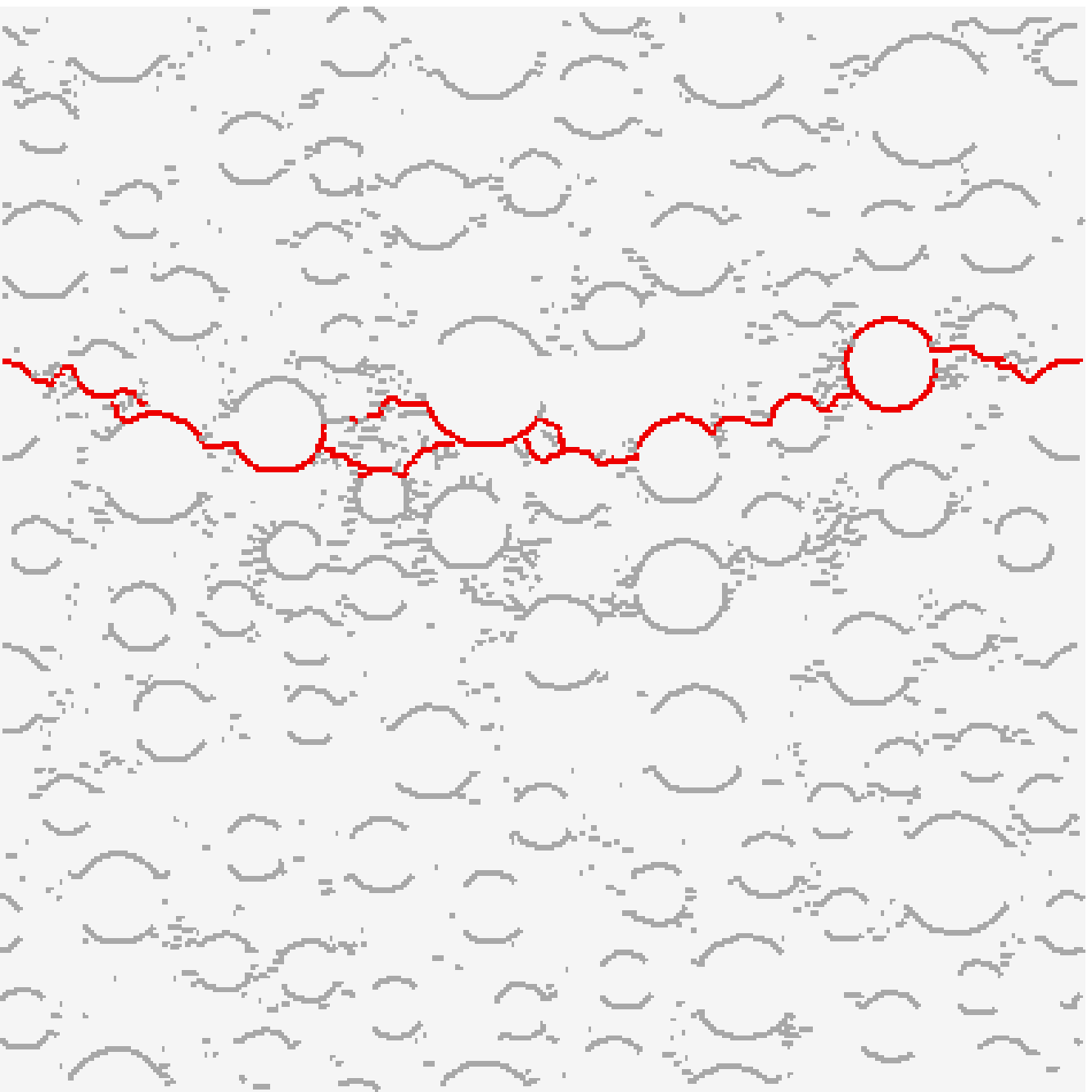,width=4cm} & \epsfig{file=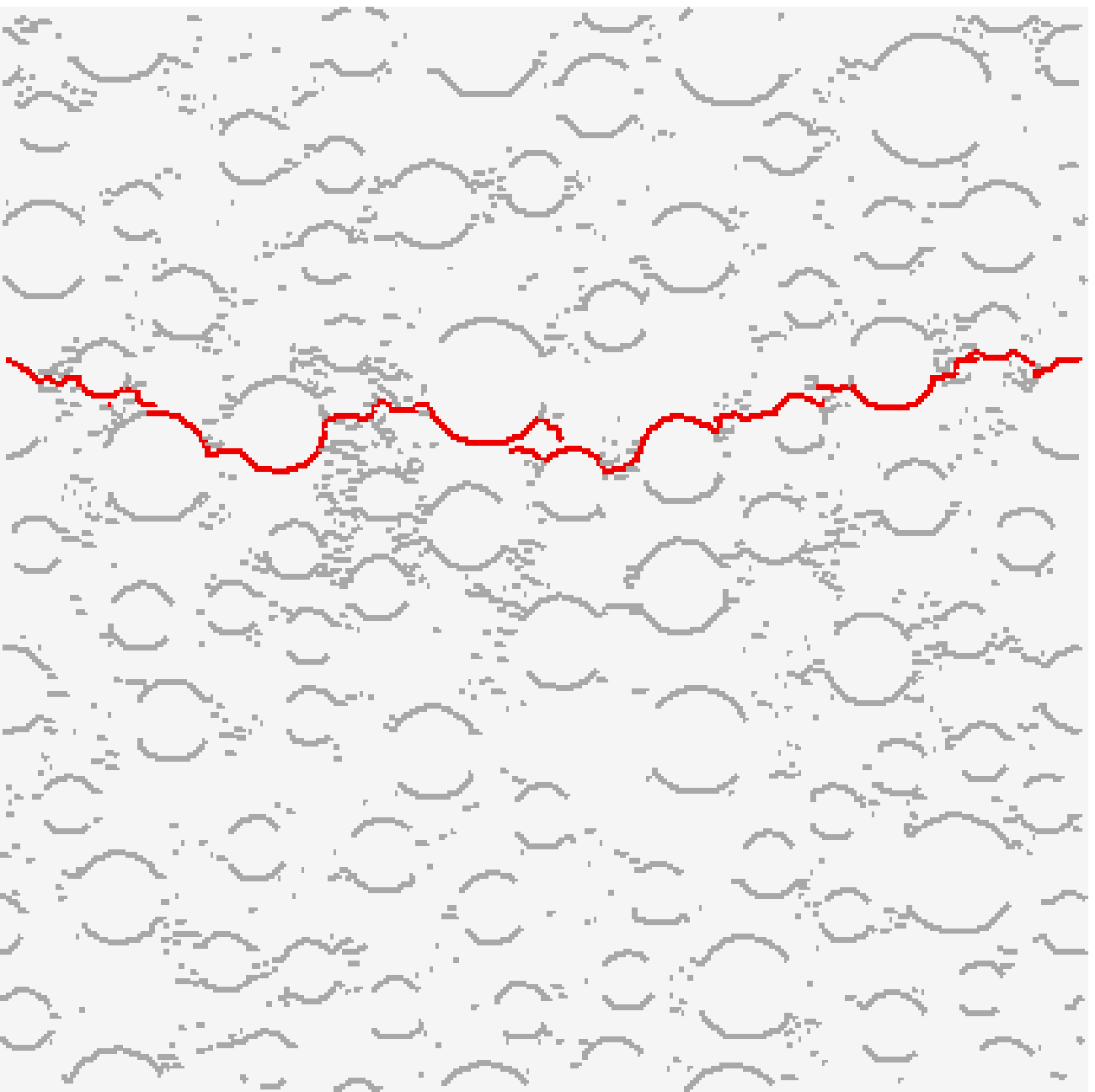,width=4cm} & \epsfig{file=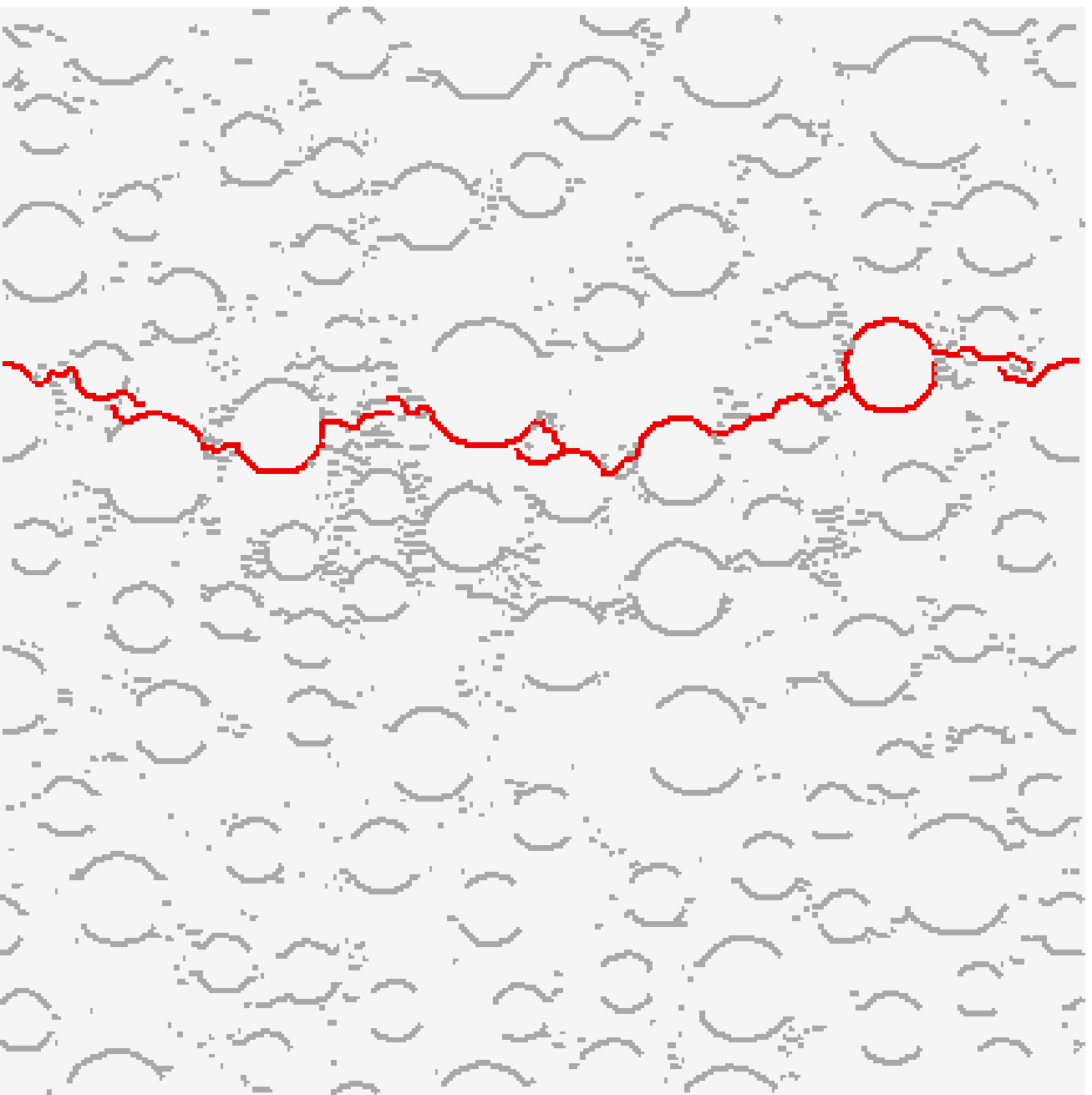,width=4cm}
\end{tabular}
\end{center}
\caption{Comparison of crack patterns obtained from analyses using three different lattices with the same meso-structure and the same average lattice element size.}
\label{fig:crackRandom}
\end{figure}

Additionally, the influence of the lattice on the results was investigated. 
In the present study, the size of lattice elements and their spatial arrangement was chosen independently of the  material meso-structure.
Of course, the lattice elements must be sufficiently small to be able to represent the discretely modelled aggregates.
However, fracture is modelled as displacement jumps within individual lattice elements.
Therefore, the spatial orientation and size of the lattice elements might influence the fracture patterns.
The influence of these two parameters is investigated by two additional studies. 
Firstly, analyses with three different random lattices with the same minimum distance $d_{\rm min}$ were conducted. 
The same aggregate arrangement and random field was used. 
The stress-strain curves and crack patterns of these analyses are compared in Fig.~\ref{fig:ldRandom}~and~\ref{fig:crackRandom}. 
The stress-strain curves exhibit only small differences in the post-peak regime, which are smaller than the standard deviations of the stress-strain curves shown in Fig.~\ref{fig:ld}.
The overall crack patterns in Fig.~\ref{fig:crackRandom}, which determine the shape of the fracture process zone in Fig.~\ref{fig:fpzAverage}, are almost independent of the background lattice. Only small differences are visible, which have a negligible influence on the tortuosity of the fracture process zones obtained.

\begin{figure}
\begin{center}
\epsfig{file=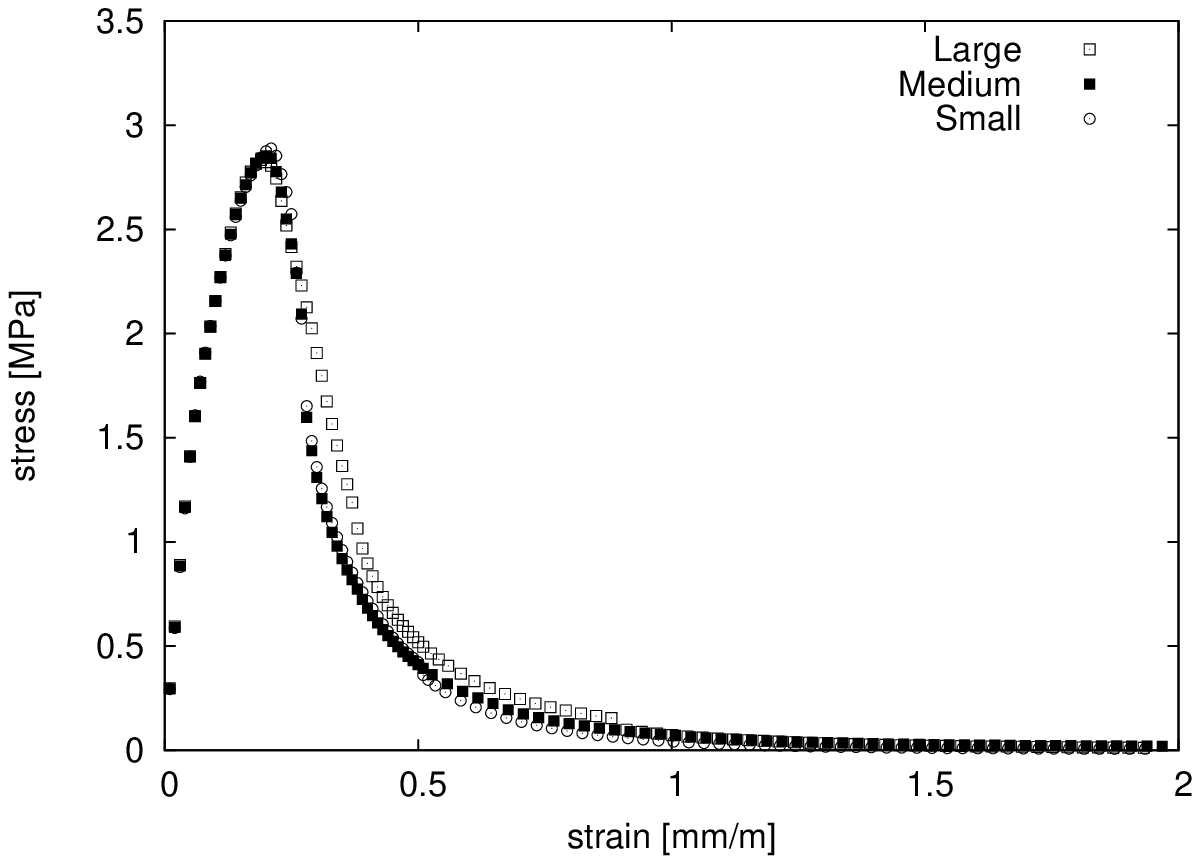,width=10cm}
\end{center}
\caption{Comparison of average stress-strain curves obtained from lattices with minimum distances of $d_{\rm min} = 1$, $0.75$~and~$0.5$~mm, respectively. In all three analyses the same aggregate arrangement and random field is used.}
\label{fig:ldSize}
\end{figure}

\begin{figure}
\begin{center}
\begin{tabular}{ccc}
\epsfig{file=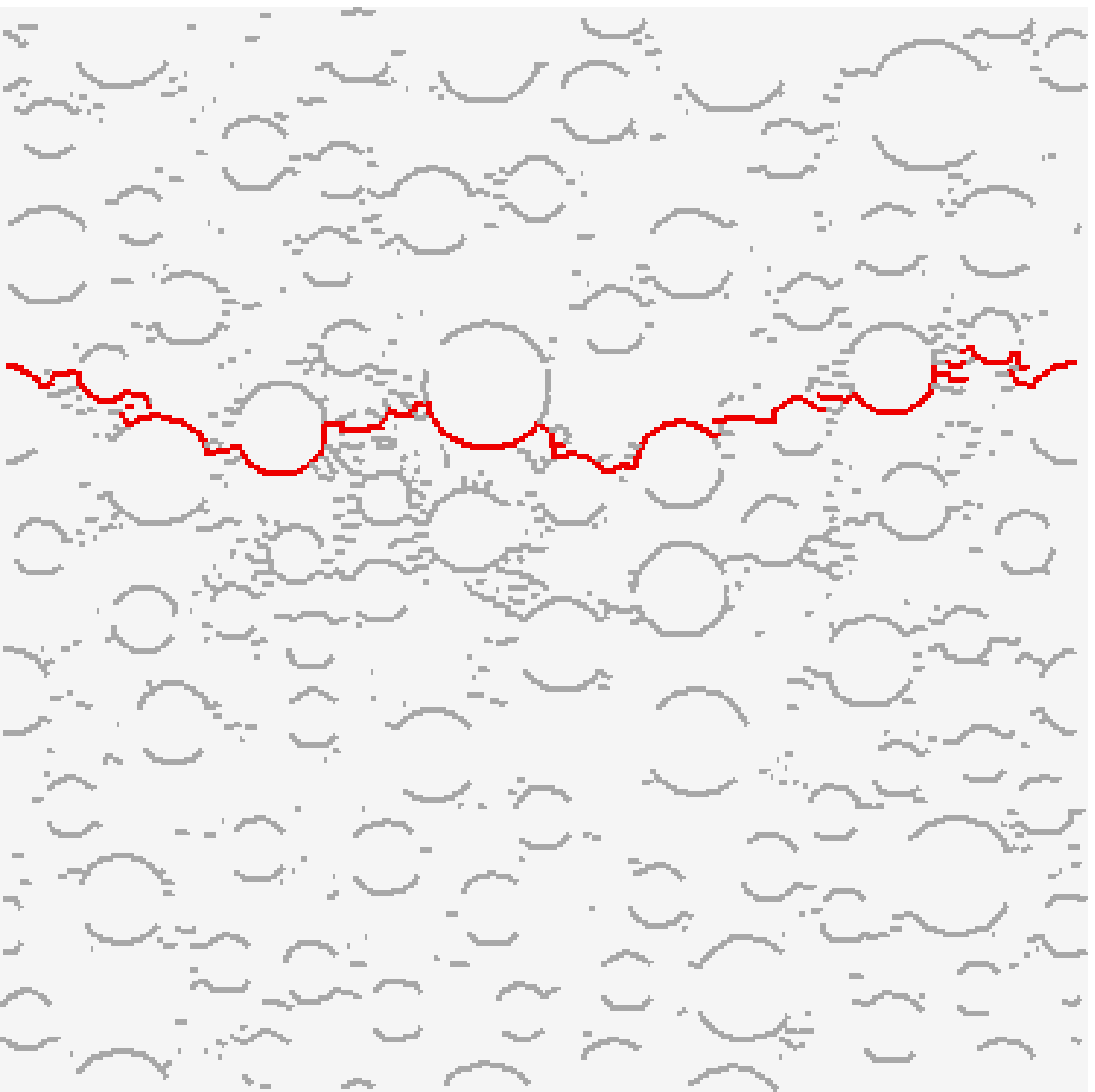,width=4cm} & \epsfig{file=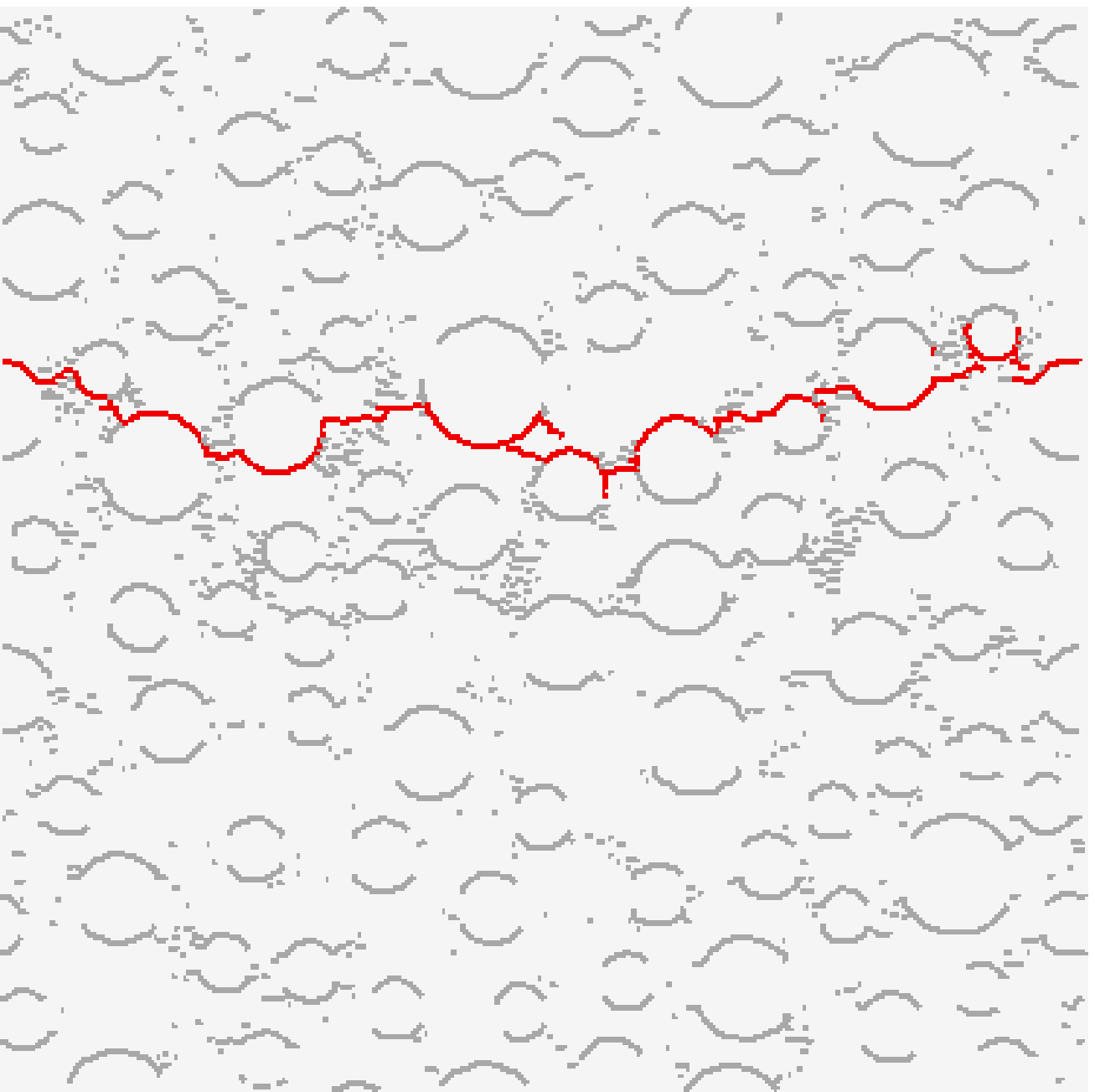,width=4cm} & \epsfig{file=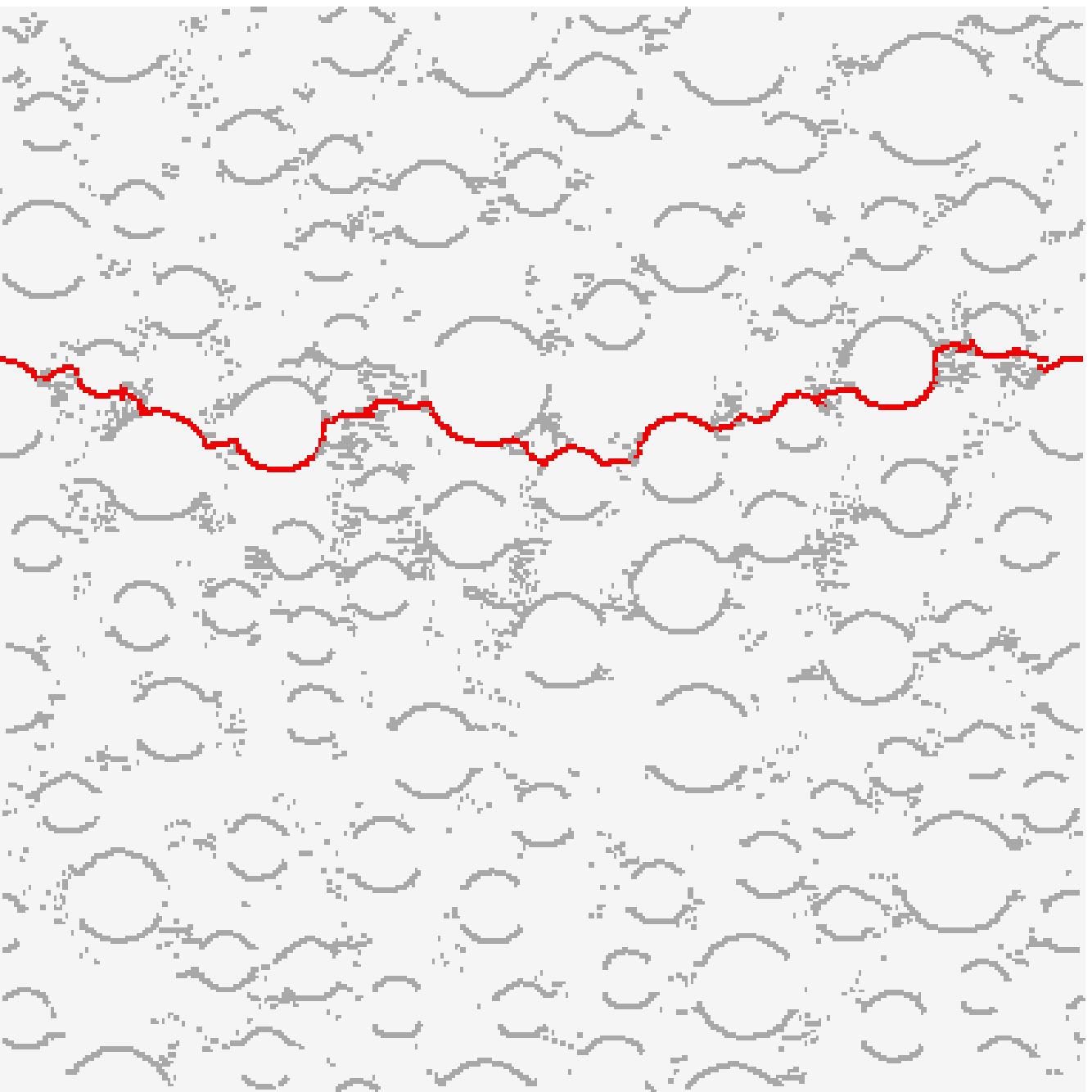,width=4cm}
\end{tabular}
\end{center}
\caption{Comparison of crack patterns obtained from lattices with minimum distances of $d_{\rm min} = 1$~mm, $d_{\rm min} = 0.75$~mm and $d_{\rm min} = 0.5$~mm, respectively. In all three analyses, the same aggregate arrangement and random field is used.}
\label{fig:crackSize}
\end{figure}

As the next step, the influence of the lattice element size was studied.
In the constitutive model used in the present study, fracture is described by a stress-crack opening curve. Therefore, the results are expected to be independent of the size of the lattice elements.
Three analyses with the same meso-structure but different background lattices with minimum distances $d_{\rm min} = 1$, $0.75$~and~$0.5$~mm, respectively, were performed. 
The differences among the stress-strain curves (Fig.~\ref{fig:ldSize}) are small, which shows that the present lattice approach is free of pathological mesh dependence.
Also, the crack patterns in Fig.~\ref{fig:crackSize} exhibit only local differences, which are due to the random arrangement of lattice elements (compare with Fig.~\ref{fig:crackRandom}), but not their size.
These local differences are not expected to influence the average fracture process zones presented in Fig.~\ref{fig:fpzAverage}.
The two studies above show that discretely modelled aggregates in connection with a random field for the mortar provide results which are almost independent of the background lattice.


\begin{figure}
\begin{center}
\epsfig{file=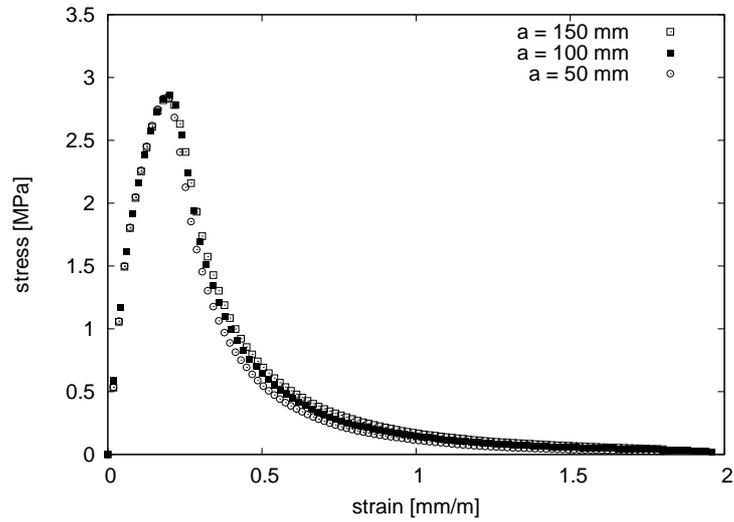,width=10cm}
\end{center}
\caption{Comparison of average stress-strain curves from 100 analyses obtained for periodic cells of size (perpendicular to the loading direction) $a = 50$, $100$ and $150$~mm.}
\label{fig:ldWidth}
\end{figure}

\begin{figure}
\begin{center}
\epsfig{file=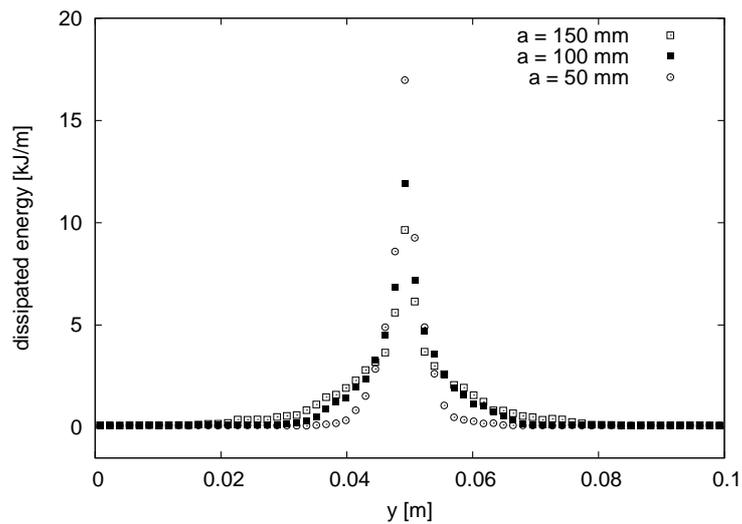,width=10cm}
\end{center}
\caption{Comparison of average fracture process zones obtained for periodic cells of sizes $a=50$, $100$ and $150$~mm. Fracture process zones are shifted before averaging.}
\label{fig:fpzWidth}
\end{figure}

A periodic cell was chosen in the present study so that the results are independent of the boundary conditions and the fracture process zones can be shifted before the averaging of the results. 
The size of the periodic cell was chosen so that the tortuosity of the fracture patterns is statistically representative. A decrease of the size of the periodic cell in the direction of loading is expected to influence the tortuosity of the crack patterns, i.e.\ the width of the fracture process zone.
The smaller the cell size, the less tortuous are the crack patterns.
However, it is expected that there is an upper limit of the cell size, above which the width of the fracture process zone remains almost constant.
The influence of the cell size was studied by performing two sets of 100 analysis for two additional cells with $a = 50$~and~$150$~mm.
The average stress-strain curves and fracture process zones for these two sizes are compared to the original results for $a = 100$~mm in Figs.~\ref{fig:ldWidth}~and~\ref{fig:fpzWidth}, respectively. 
Although the difference between the average stress-strain curves for the three sizes is small, it can be seen that a reduction of the size leads to a steeper softening branch in the post-peak regime of the stress-strain curves.
Furthermore, the comparison of the fracture process zones shows that the width of the fracture process zone reduces considerably when the size of the periodic cell is decreased from $100$~to~$50$~mm.
On the other hand, an increase  of the size from $100$ to $150$~mm has only a small influence on the width of the fracture process zone. 
Consequently, a cell size of $100$ mm was deemed to be sufficient to represent the fracture process zone. 

\section{Comparison of meso-scale analyses to macroscopic nonlocal constitutive model}

In the previous two sections, the meso-scale description of fracture in concrete was used to determine the average of dissipated energy densities. 
The present study adopts the view that this average of densities corresponds to the fracture process zone, which is represented by macroscopic nonlocal models.
In the following section, a one-dimensional macroscopic nonlocal damage model is compared to the  results of meso-scale analyses.

\subsection{Nonlocal damage model}

In the one-dimensional setting, the stress-strain law used by the damage model is
\begin{equation}
\sigma = \left( 1 - \omega \right) E \varepsilon
\end{equation}
where $\sigma$ is the uniaxial stress, $\omega$ is the damage variable, $E$ is Young's modulus and $\varepsilon$ is the strain.
Damage evolution is driven by the nonlocal strain $\bar{\varepsilon}$,
which represents a weighted spatial average of the local strain $\varepsilon$
(in multiple dimensions, a scalar measure of the strain level called the equivalent strain would be used).
In an unbounded one-dimensional medium, the nonlocal strain is evaluated as
\begin{equation}
\bar{\varepsilon}(x) = \int_{-\infty}^{\infty} \alpha (x-\xi) \varepsilon (\xi)\, {\rm d} \xi
\end{equation}
where $\alpha$ is the nonlocal weight function, decaying with increasing distance between points
$x$ and $\xi$ and normalized such that the averaging operator does not modify a uniform field.
The weight function is sometimes taken as a Gauss-type function,
but in the present study we consider 
the quartic polynomial with a bounded support, 
\begin{equation}\label{quartic}
\alpha(r) = \frac{15}{16\,R} \left\langle 1-\frac{r^2}{R^2}\right\rangle^2
\end{equation}
and the exponential function 
\begin{equation}\label{green}
\alpha(r) = \frac{1}{2\,l} \exp\left(-\frac{\vert r\vert}{l}\right)
\end{equation}
Both weight functions are normalized such that $\int_{-\infty}^{\infty} \alpha (r)\, {\rm d} r=1$.
Parameters $R$ or $l$ reflect the characteristic length of the material and have a strong influence
on the width of the process zone, as will be shown later.
Function (\ref{green})
corresponds to Green's function of the differential equation 
$\bar{\varepsilon}-l^2\bar{\varepsilon}''= \varepsilon$
(on an infinite one-dimensional domain). 
With this specific choice,
the nonlocal integral-type model is equivalent to the implicit gradient model proposed in \citep{Peerlings96}. 

Since damage is irreversible,
the damage variable $\omega$ is related to  an internal variable $\kappa$,
which represents the maximum level of nonlocal strain ever reached
in the previous history of the material. Formally, $\kappa$ can be defined by the 
loading-unloading conditions (\ref{loadunload}) with the loading function
\begin{equation}
f(\bar{\varepsilon},\kappa)=\bar{\varepsilon}-\kappa
\end{equation}
The damage law that links the strain to damage is closely related to the shape of the 
stress-strain curve. Since the curves obtained from the lattice model exhibit nonlinearity
already before the peak, we consider a damage law that can reproduce such behavior:
\begin{equation} \label{eq:damageEvol}
\omega = g(\kappa) = \left \{ \begin{array}{ll} 
1-\exp\left(-\displaystyle\frac{1}{m}\left(\dfrac{\kappa}{\varepsilon_{\rm p}}\right)^m\right) & \mbox{if $\kappa \leq \varepsilon_1$}\\[2mm]
1- \dfrac{\varepsilon_3}{\kappa} \exp \left(- \dfrac{\kappa- \varepsilon_1}{\varepsilon_{\rm f} \left[1+ \left(\frac{\kappa-\varepsilon_1}{\varepsilon_2}\right)^n\right]}\right) & \mbox{if $\kappa > \varepsilon_1$} \end{array} \right.
\end{equation}
The primary model parameters are the uniaxial tensile strength $f_{\rm t}$,
the strain at peak stress (under uniaxial tension) $\varepsilon_{\rm p}$, and additional parameters
$\varepsilon_1$, $\varepsilon_2$ and $n$, which control the post-peak part of the stress-strain law.
Other parameters that appear in (\ref{eq:damageEvol}) can be derived from the condition of zero
slope of the stress-strain curve at $\kappa=\varepsilon_{\rm p}$ and from the conditions of stress and stiffness continuity at $\kappa=\varepsilon_1$:
\begin{eqnarray}\label{eqparams1}
m&=&\frac{1}{\ln(E\varepsilon_{\rm p}/f_{\rm t})}\\
\label{eqparams2}
\varepsilon_{\rm f}&=&\frac{\varepsilon_1}{\left(\varepsilon_1/\varepsilon_{\rm p}\right)^m -1}\\
\varepsilon_3&=&\varepsilon_1\exp\left(-\frac{1}{m}\left(\frac{\varepsilon_1}{\varepsilon_{\rm p}}\right)^m\right)
\label{eqparams3}
\end{eqnarray}

\subsection{Parameter identification}

The parameters have been adjusted so as to fit the uniaxial stress-strain curve in Fig.~\ref{fig:ld} and the fracture process zone in  Fig.~\ref{fig:fpzAverage}. The basic parameters $E=29.6$ GPa, $f_{\rm t}=2.86$ MPa and $\varepsilon_{\rm p}=0.198\times 10^{-3}$ can be directly determined from the ascending branch of the stress-strain curve,
and the corresponding value of $m=1.39$ is calculated from (\ref{eqparams1}).
For the one-dimensional model, the strain profile remains uniform up to the
peak, and so the above-mentioned parameters are independent of the characteristic length of the nonlocal model.
The remaining parameters are related to the descending branch and their
values must be optimized for each specific choice of the nonlocal weight function separately. Here we consider two different nonlocal weight functions,
(\ref{quartic}) and (\ref{green}), each with three different values of the characteristic
length parameter $R$ or $l$. The resulting six sets of parameters are listed in
Table~\ref{tab:parameters}. Parameter $n$ controls the shape of the tail
of the stress-strain curve and can be taken by the same value $n=0.85$ for
all the parameter sets considered here. Parameters   
$\varepsilon_1$ and $\varepsilon_2$ were obtained by fitting and
parameters $\varepsilon_{\rm f}$ and $\varepsilon_3$ were then calculated from
(\ref{eqparams2})--(\ref{eqparams3}).

\begin{table}
\caption{Parameters that control the post-peak part of the stress-strain curve.}
\label{tab:parameters}
\centering
\begin{tabular}{|ccc|cc|cc|}
\hline
parameter & weight  & length  & $\varepsilon_1$ & $\varepsilon_2$ & $\varepsilon_3$ & $\varepsilon_{\rm f}$
\\
set & function & parameter & [$10^{-3}$] & [$10^{-3}$] & [$10^{-3}$] & [$10^{-3}$]
\\
\hline
A & (\ref{quartic}) & $R=20$ mm & 0.255 & 5 & 0.0919 & 0.603 
\\
B & (\ref{quartic}) & $R=15$ mm & 0.238 & 7 & 0.0942 & 0.814 
\\
C & (\ref{quartic}) & $R=10$ mm & 0.221 & 15 & 0.0958 &  1.335
\\
D & (\ref{green}) & $l=8$ mm & 0.280 & 2.5 & 0.0875 & 0.451
\\
E & (\ref{green}) & $l=6$ mm & 0.255 & 3.5 & 0.0919 & 0.603
\\
F & (\ref{green}) & $l=4$ mm & 0.230 & 7 & 0.0950 & 0.990
\\
\hline
\end{tabular}
\end{table}

\begin{figure}
\begin{center}
\epsfig{file=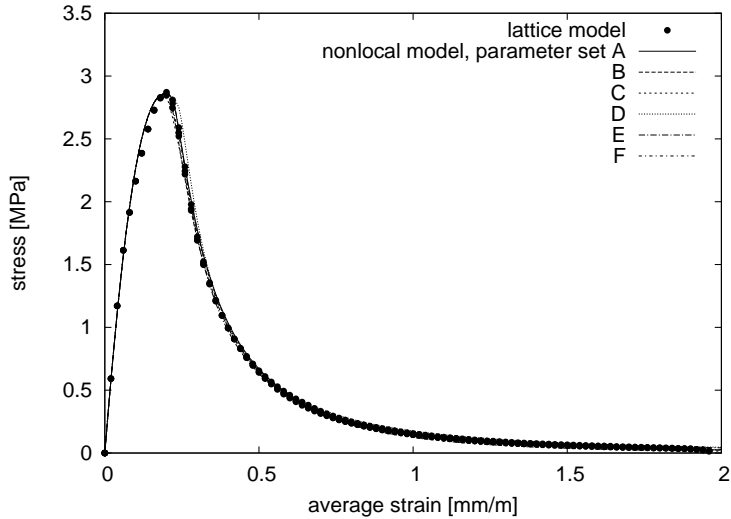,width=10cm}
\end{center}
\caption{Comparison of average stress-strain curves obtained with the meso-scale lattice model and with the macroscopic nonlocal model using different combinations of parameters.}
\label{fig:ldComp}
\end{figure}

\begin{figure}
\centering
\begin{tabular}{c}
(a) 
\\
\epsfig{file=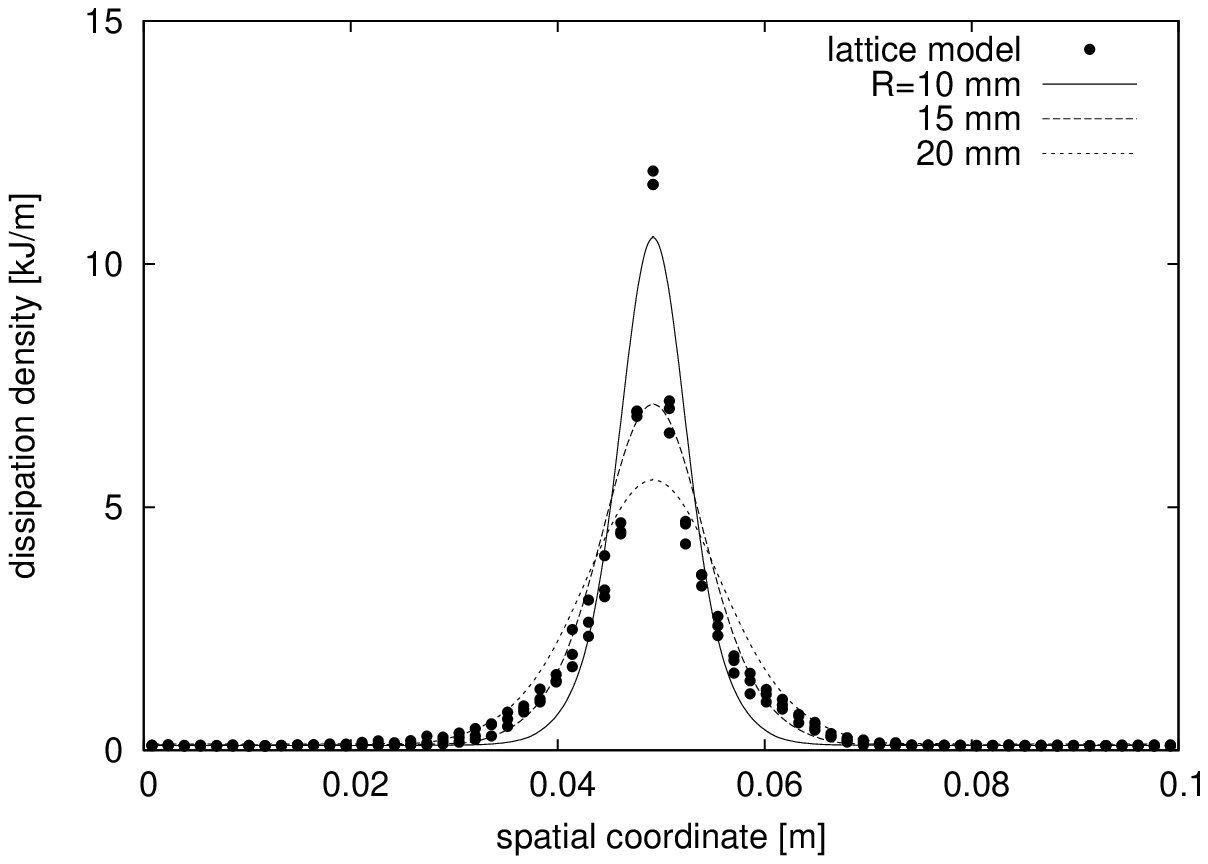,width=10cm}
\\[4mm]
(b)
\\
\epsfig{file=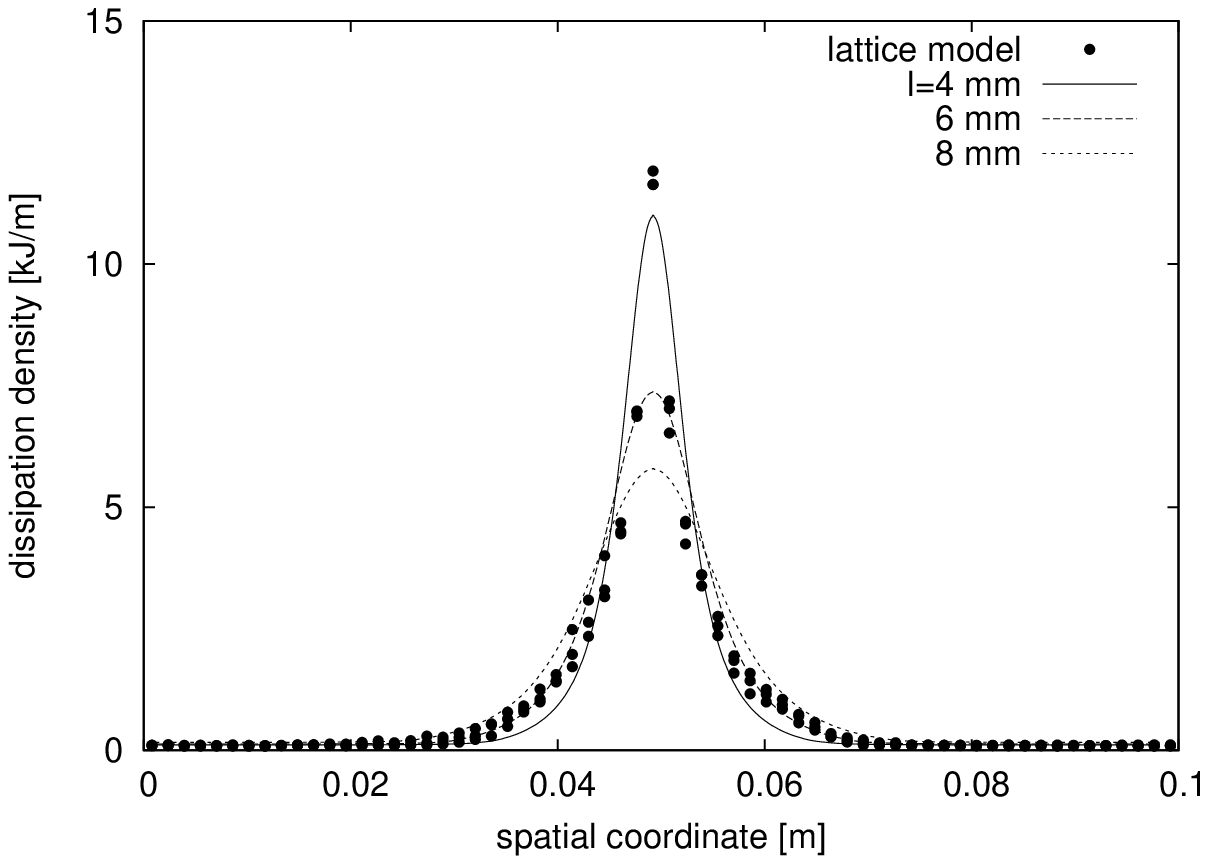,width=10cm}
\end{tabular}
\caption{Distributions of the dissipated energy density across the fracture process zone obtained with the meso-scale lattice model and with the macroscopic nonlocal model using (a) quartic polynomial weight function (\ref{quartic}), (b) exponential (Green-type) weight function (\ref{green})}
\label{fig:fpzComp}
\end{figure}

The comparison of the nonlocal model and the meso-scale approach is shown in Figs.~\ref{fig:ldComp} and \ref{fig:fpzComp}.
With a proper choice of parameters of the macroscopic nonlocal model, 
the damage evolution law (\ref{eq:damageEvol}) allows for a good fit of the average stress-strain curve obtained from the meso-scale lattice analyses;
see Fig.~\ref{fig:ldComp}. The agreement is almost perfect for all six
sets of parameters presented in Table~\ref{tab:parameters}. This means that 
the same global response can be reproduced with different types of nonlocal
weight functions and different values of the length parameter $R$ or $l$.
The optimal value of the characteristic length cannot be deduced from 
the average stress-strain curve, which is in fact the rescaled load-displacement
curve and characterizes the global response. 
As shown in Fig.~\ref{fig:fpzComp}, the characteristic length
controls the width of the process zone. 
\begin{figure}
\centering
\epsfig{file=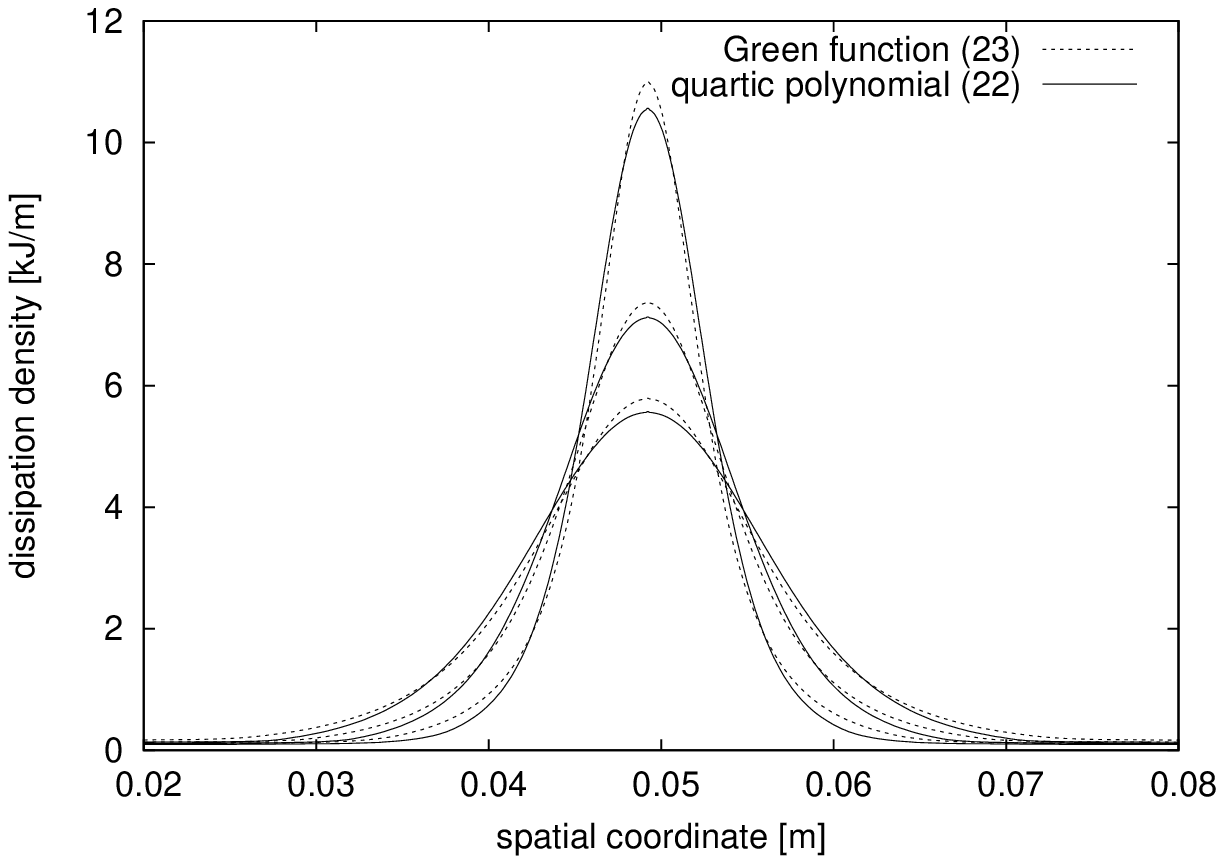,width=10cm}
\caption{Distributions of the dissipated energy density across the fracture process zone obtained with the macroscopic nonlocal model using quartic polynomial weight function (\ref{quartic}) and exponential (Green-type) weight function (\ref{green})}
\label{fig:fpzComp2}
\end{figure}

\begin{figure}
\centering
\epsfig{file=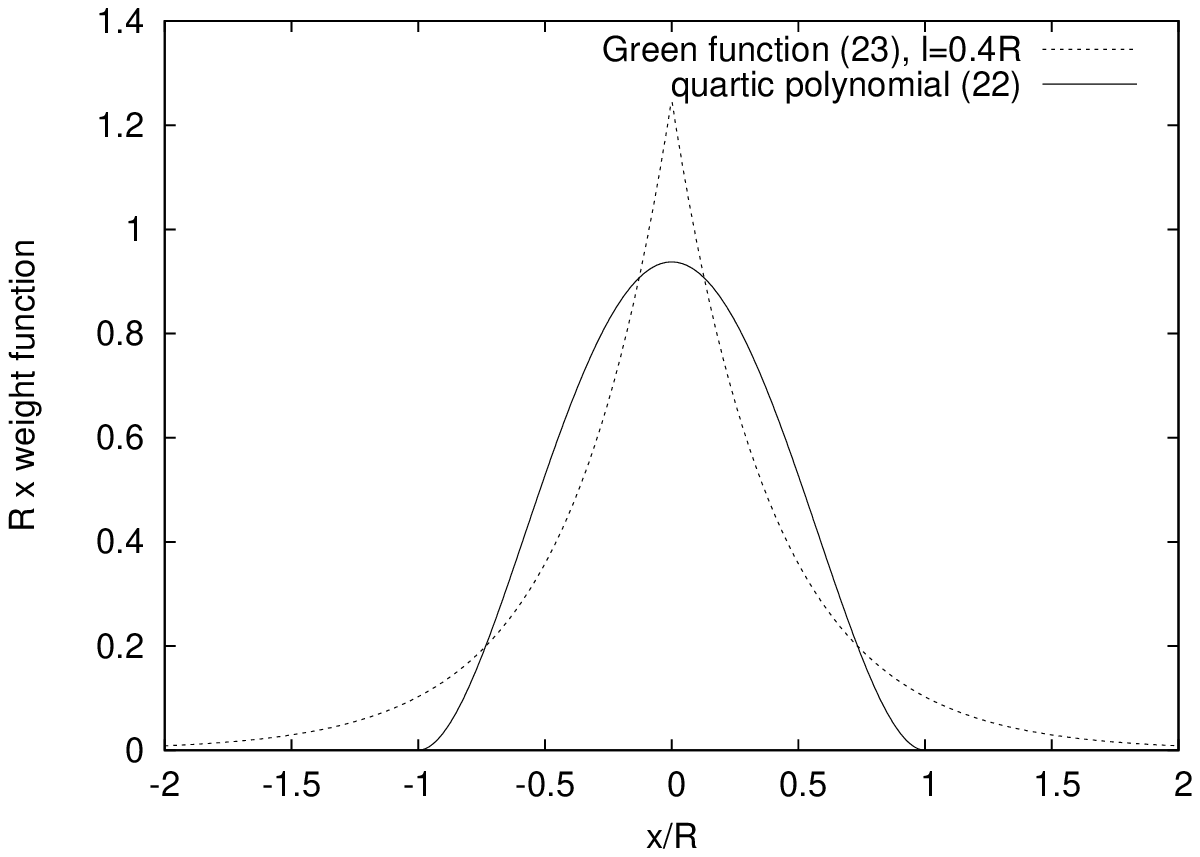,width=10cm}
\caption{Quartic polynomial weight function (\ref{quartic}) and exponential (Green-type) weight function (\ref{green}) with the ratio of length parameters $R/l=2.5$.}
\label{fig:fpzComp3}
\end{figure}

On the other hand, the specific form of the weight function 
has only a weak influence on the distribution of energy dissipation across
the process zone; see Fig.~\ref{fig:fpzComp2}. The exponential weight function
gives a slightly higher peak of the dissipation profile but this difference
is much less pronounced than the difference in the weight functions;
see Fig.~\ref{fig:fpzComp3}. Of course, the length parameters $R$ and $l$
do not have the same meaning; similar dissipation profiles are obtained
if $R$ is set approximately to $2.5\,l$.
For  $R=15$ mm or $l=6$ mm, a good fit of the dissipation profile predicted
by the meso-mechanical lattice model is achieved, except for the center
of the process zone where the extremely high dissipation density obtained
with the lattice model is underestimated.

\section{Conclusions}
In the present work, a meso-scale approach was used to determine the fracture process zone of concrete, which is characterized by the average of dissipated energy densities. 
Then, a macroscopic nonlocal model was fitted to the meso-scale results.
The work resulted in the following conclusions:
\begin{itemize}
\item The average of dissipated energy densities obtained from the meso-scale analyses resulted in a fracture process zone of a finite width, which is determined by the tortuosity of the crack path.
\item The meso-scale results obtained with a lattice approach are independent of the size and spatial orientation of the elements. This is achieved by describing the material properties by a random field of strength. 
\item The standard deviation of the energy densities obtained is greater than the mean value, which indicates a deviation from the normal distribution. Nevertheless, the results were shown to be statistical representative.
\end{itemize}
In future work, the meso-scale results will be used to calibrate nonlocal macroscopic damage models reported in the literature, which describe boundaries differently.
These nonlocal models will then be applied to the analysis of notched concrete beams, for which boundary conditions play an important role.
Additionally, meso-scale analysis of these concrete beams will be carried out, which will allow the identification of nonlocal models suitable for the description of the failure of concrete.

\section*{Acknowledgements}

Financial support of the Ministry of Education, Youth and Sports of the Czech Republic under Research Plan MSM6840770003 is gratefully acknowledged by the second author. The authors would also like to express their gratitude to Dr. Bo\v{r}ek Patz\'{a}k of the Czech Technical University for kind assistance with his finite element package OOFEM \citep{Pat99,PatBit01}.

\bibliographystyle{plainnat}

\bibliography{general}

\end{document}